\documentclass[aps,11pt,showkeys,nofootinbib]{revtex4-2}

\usepackage[margin=1in,letterpaper]{geometry}
\usepackage{amsmath,amssymb,amsfonts,amsthm}
\usepackage{physics}
\usepackage{graphicx}
\usepackage{hyperref}
\usepackage{cleveref}

\usepackage{diagbox}
\usepackage[table,xcdraw]{xcolor}

\usepackage{tikz}
\usetikzlibrary{decorations.pathmorphing,patterns}

\usepackage{subcaption}
\usepackage{wrapfig}

\newcommand{\td}{\text{d}}

\begin{document}
\title{\bf Unimodular boundary time for Regge calculus}

\author{Steffen Gielen, Sofie Ried}
\affiliation{School of Mathematical and Physical Sciences, University of Sheffield, Hicks Building, Hounsfield Road, Sheffield S3 7RH, United Kingdom}
\email{s.c.gielen@sheffield.ac.uk}
\email{sried1@sheffield.ac.uk}

\begin{abstract}
Henneaux--Teitelboim gravity is a diffeomorphism-invariant formulation of unimodular gravity that is locally equivalent to general relativity but includes an extra boundary variable, unimodular time, whose difference between hypersurfaces measures the enclosed 4-volume. In general, unimodular versions of quantum gravity can be expected to deviate from theories based on general relativity, but a discrete path-integral formulation for Henneaux--Teitelboim gravity has not yet been established. In this paper, we discretise Henneaux--Teitelboim gravity using the methods of Euclidean Regge calculus. Testing this framework in a symmetry-reduced model of Regge cosmology, we reproduce results in previous literature and demonstrate that the time continuum limit can be naturally defined using unimodular time rather than proper time. Because unimodular time appears as boundary data as opposed to being a degree of freedom in the bulk, this approach allows for the comparison of general simplicial triangulations, beyond highly symmetric discretisations, to the continuum. Our work provides a first step towards a discrete path-integral formulation of Henneaux--Teitelboim gravity and adds a new perspective on time as a boundary variable in quantum gravity.
\end{abstract}


\maketitle

\section{Introduction}

Regge calculus is a theory of discrete gravity based on a discrete analogue of the Einstein--Hilbert action of general relativity. It was first introduced by Tullio Regge in his seminal 1961 paper ``General relativity without coordinates" \cite{Regge:1961px}. Instead of a metric, which depends on coordinates, the geometry is defined in terms of edge lengths of a triangulation of a manifold. Regge calculus was originally formulated in Euclidean signature, and while there is literature on Lorentzian Regge calculus \cite{Tate:2011ct,Sorkin:2019llw,Ito:2022ycc,Dittrich:2021gww,Asante:2021phx,Borissova:2023izx}, in this paper we will stick with the original Euclidean signature for simplicity. Regge calculus can be used to approximate continuum spacetimes \cite{Rocek:1981ama,Rocek:1982tj,Barrett:1988wd,Wong:1971,Gentle:1998,Brewin:2000zh}, but it has also been influential in the development of several theories of quantum gravity. The initial idea of quantum Regge calculus is to define a path integral for discrete quantum gravity directly from the dynamical variables and action of Regge calculus; this leads to an object that is in some sense better behaved than the continuum path integral \cite{Hartle:1985wr}. However, it still inherits some pathologies of the continuum path integral, and Euclidean dynamical triangulations (EDT) and finally causal dynamical triangulations (CDT) were developed to rid the path integral of seemingly unphysical behaviour \cite{Ambjorn:2005jj,Loll:2019rdj}. The spin foam approach and the group field theory approach \cite{Rovelli:2014ssa,Oriti:2007qd} are also related to quantum Regge calculus, although a little less directly.

In this paper we want to apply a discretisation à la Regge to the Henneaux--Teitelboim formulation of unimodular gravity \cite{henneaux_cosmological_1989}. Unimodular gravity is classically equivalent to general relativity, apart from a different treatment of the cosmological constant as an integration constant rather than a fundamental constant of Nature. One motivation for unimodular gravity is to define a more minimalistic version of general relativity, removing some redundancy in the theory: the trace part of the Einstein equations does not contain any physical information beyond the value of the cosmological constant, and can almost entirely be recovered from the trace-free part of the Einstein equations and the Bianchi identities \cite{Ellis:2010uc}. In unimodular gravity, the equations of motion are given only by the trace-free Einstein equations. Apart from the appeal of being more minimalistic than general relativity, this reformulation also gives a different perspective on the cosmological constant as an initial condition rather than a fundamental parameter in the action.  Because the cosmological constant is treated differently, it is reasonable to assume that the quantisation of general relativity and unimodular gravity would be different, since the latter would allow for superpositions of different values of the cosmological constant. The questions of equivalence at the quantum level and of implications for the so-called cosmological constant problem have been addressed by many authors (see, e.g., \cite{Smolin:2009ti,deLeonArdon:2017qzg,Percacci:2017fsy,Jirousek:2023gzr}). When matter is included, the discussion is a bit more subtle, since energy-momentum conservation no longer follows from the Bianchi identities and can in principle be violated, leading to potentially interesting phenomenology \cite{Ellis:2010uc,Josset:2016vrq,Perez:2017krv}. In this paper we are only interested in vacuum gravity.

So far there is not much literature on fundamental quantum gravity theories based on unimodular gravity (approaches that have been discussed include unimodular loop quantum gravity and asymptotic safety \cite{Smolin:2010iq,Eichhorn:2013xr,Eichhorn:2015bna,Benedetti:2015zsw,DeBrito:2019gdd,deBrito:2020rwu,deBrito:2020xhy}). Since many approaches to quantum gravity in one form or another are related to Regge calculus, Regge calculus provides a good starting point for a  definition of a fundamental (discrete) unimodular quantum gravity theory. Concretely, we will be working with the diffeomorphism-invariant Henneaux--Teitelboim formulation of unimodular gravity. In this formulation, there is an additional degree of freedom called unimodular time, which is related to the 4-volume between hypersurfaces. This notion of time between hypersurfaces helps connect the discrete solutions of Regge calculus to the continuum, as we will demonstrate in \cref{sec:models}. Since unimodular time is a boundary variable, a quantisation would potentially promote it to an operator on a boundary Hilbert space, making the analysis of time in quantum gravity accessible in a way that is different from non-unimodular formalisms. 

In general the equations of motion of Regge calculus are highly nontrivial to solve, as they are composed of coupled nonlinear differential equations. The study of spacetimes with a high amount of symmetry is nonetheless possible; cosmological Regge models of the evolution of a 3-sphere have been discussed in the literature \cite{Collins:1973mh,Liu:2015gpa,Dittrich:2021gww}. We will use such cosmological models to compare our theory of unimodular Regge calculus with standard Regge calculus, reproducing some previous results. We will also show that unimodular time provides a way to take the time continuum limit of the Regge action using the edge lengths as variables (as in Regge's original formulation), without having to reformulate it in terms of the height of 4-dimensional building blocks. Building on this we will then also consider a less symmetric (but still symmetry reduced) and completely simplicial discretisation of the bulk between two boundary 3-spheres. Using unimodular time in this discrete setting we will compare dynamical solutions of Regge calculus to continuum solutions, even though in this case the bulk geometry does not allow for a time continuum limit and there is no natural proxy for proper time. We show that for $\Lambda > 0$ this simplicial model agrees better with the continuum than the more symmetrical one discussed earlier. This demonstrates the advantages of including unimodular time into Regge calculus already at the classical level. 

The structure of the paper is as follows. We introduce the continuum Henneaux--Teitelboim theory of unimodular gravity in \cref{sec:continuum}, and specify to cosmological solutions in \cref{minisupersec}. We introduce the standard formulation of Regge calculus in \cref{sec:regge}, which allows us to define the new Henneaux--Teitelboim--Regge action in \cref{sec:uniregge}. In \cref{sec:cosmomodel} we compare this formalism to standard Regge calculus via the cosmologial model of the evolution of a discretised 3-sphere; we then introduce a different discretisation in \cref{sec:simmodel} which is shown to have better agreement with the continuum for $\Lambda > 0$, a comparison for which we make use of the new boundary variable of unimodular time. We conclude and remark on possible future extensions of this research in \cref{sec:conclusion}. Unless otherwise stated, we will work in units where $c = \hbar = 8\pi G = 1$.

\section{Unimodular Gravity in the Continuum}
\label{sec:continuum}

The Einstein equations of general relativity have a redundancy in them: the trace part can be inferred from the trace-free equations and the Bianchi identity $\nabla_{\mu}R^{\mu}_{\,\,\nu} = \frac{1}{2}\nabla_{\nu}R$ up to an integration constant \cite{Ellis:2010uc}. This integration constant gives an effective cosmological constant. This was already noticed by Einstein in 1919 \cite{Einstein1919} (a translation of this paper can be found in \cite{lorentz1952principle}). Already at the linearised level, one can choose perturbations around a background metric to be trace-free, as the trace does not contain any physical degrees of freedom. To get rid of this redundancy one can fix the metric determinant, e.g., choose $|g| = 1$, and remove the trace part of variations, meaning that only the trace-free part of the Einstein equations is obtained. Theories that incorporate these principles at the level of the action or field equations go under the name of unimodular gravity \cite{Anderson:1971pn,vanderBij:1981ym,Buchmuller:1988wx,Unruh:1988in,Sorkin:1987cd,gauging}. 

Up to the difference in interpreting the cosmological constant, unimodular gravity is equivalent to general relativity (and so in particular satisfies all observational tests of general relativity), but it is not fully diffeomorphism-invariant. Since we have fixed the determinant of the metric, the theory is only invariant with respect to volume-preserving diffeomorphisms. One way to restore full diffeomorphism invariance is to introduce new variables $T^{\mu}$ and fix the metric determinant by
\begin{align}
\label{eq:unicon}
    \sqrt{|g|} = \partial_{\mu}T^{\mu}\,.
\end{align}
This is the approach taken in the Henneaux--Teitelboim formalism \cite{henneaux_cosmological_1989}, which we will introduce in more detail now. For a recent review highlighting various particular perspectives on unimodular gravity, see for example \cite{carballo-rubio_unimodular_2022}.

\subsection{The Henneaux--Teitelboim Action and Unimodular Time}

The action of Henneaux--Teitelboim gravity is given by \cite{henneaux_cosmological_1989}
\begin{align}
    S_{{\rm HT}}+S_{{\rm GHY}} = \frac{1}{16\pi G}\int_M\td^4x\,\left[\sqrt{|g|}R-2\Lambda(\sqrt{|g|}-\partial_{\mu}T^{\mu})\right]+\frac{1}{8\pi G}\int_{\partial M}\td^3x\,\varepsilon\sqrt{|h|}K\,.
    \label{eq:HTaction}
\end{align}
The first two terms in the bulk Lagrangian correspond to the Einstein--Hilbert Lagrangian, where $\Lambda$ is not the cosmological constant but a Lagrange multiplier field enforcing the unimodular condition (\ref{eq:unicon}). The last term is the Gibbons--Hawking--York boundary term, which depends on the induced 3-metric $h_{ab}$ on the boundary via its determinant $h$ and the extrinsic curvature scalar $K$. Here $\varepsilon = -1$ if the normal to the boundary is timelike and $\varepsilon = 1$ if it is spacelike. We are writing $|g|$ and not $-g$, as we want this discussion to apply to both the Euclidean and Lorentzian formulations. In the transition to the discrete setting of Regge calculus we will focus on Euclidean gravity, as Lorentzian Regge calculus includes significant additional challenges \cite{Sorkin:2019llw,Dittrich:2021gww,Asante:2021phx}. With our convention $8\pi G=1$, the numerical coefficients in front of the bulk and boundary action reduce to $\frac{1}{2}$ and $1$. Notice that variation with respect to $T^\mu$ gives the dynamical equation $\partial_\mu\Lambda=0$, so that $\Lambda$ is a constant.

It should be noted that since the action only depends on $T^{\mu}$ through its divergence, we have the freedom of adding an arbitrary divergence-free vector to $T^{\mu}$,
\begin{equation}
    T^\mu \rightarrow T^\mu+\omega^\mu\,,\quad \partial_\mu\omega^\mu=0\,.
\end{equation}
Thus instead of four new degrees of freedom, we really only have one, the other three being gauge. Given $\tilde{T}^{\mu}$ one can for example try to find a $\omega^{\mu}$ such that $T^i = \tilde{T}^i+\omega^i = 0$ for three out of four components. The divergence-free condition then implies $\partial_0\omega^0=\partial_i\tilde{T}^i$, which can be solved at least locally to find $\omega^0(t,\vec{x}) = \int_0^t\text{d}t'\partial_i\tilde{T}^i(t')$.\footnote{We assume that even in Euclidean signature coordinate indices take the values $\mu=0,1,2,3$.} In regions where this solution remains well-defined, we may then gauge-fix $T^i = 0$ for $i = 1,2,3$. However, if the $t$ coordinate is periodic (e.g., because it parametrises a compact dimension), such a gauge fixing is generally not possible.

From the new degree of freedom we can construct a quantity that we may call unimodular time. Let us assume a foliation of our spacetime into hypersurfaces $\Sigma_i$. Unimodular time is then defined by contracting the oriented normal $n_{\mu}$ of the hypersurface with $T^{\mu}$ and integrating over $\Sigma_i$:
\begin{align}
    T_i = \int_{\Sigma_i}\td^3x\,n_{\mu}T^{\mu}\,.
    \label{Tidef}
\end{align}
Unimodular time is also called volume time. The reason is that if we calculate the 4-volume of a region $U$ with boundary $\partial U = \Sigma_{i+1}\sqcup\Sigma_i$ given by the union of two hypersurfaces, where the globally defined normal is pointing outward at $\Sigma_{i+1}$ and inward at $\Sigma_i$, we get the difference in unimodular time\footnote{This is under the assumption that the 4-volume between the hypersurfaces is finite.}:
\begin{align}
    V_U =\int_U \td^4 x\,\sqrt{|g|} = \int_U\td^4x\,\partial_{\mu}T^{\mu} = \int_{\Sigma_{i+1}}\td^3x\,n_{\mu}T^{\mu}-\int_{\Sigma_{i}}\td^3x\,n_{\mu}T^{\mu} = T_{i+1}-T_i\,.
\end{align}
For Lorentzian classical gravity this gives us a notion of time between two spatial hypersurfaces that is completely coordinate-independent. If the hypersurfaces are not spacelike or the metric is  Euclidean, the construction still works, but the name unimodular time may be misleading. In such cases $T_i$ does not measure time, but can still be seen as a preferred evolution parameter (see for instance \cite{QBlackHole} for an application to timelike foliations of an exterior black hole spacetime). In any case, the definition of unimodular time presupposes a fixed choice of foliation \cite{kuchar_does_1991}. 

We have introduced unimodular gravity and Henneaux--Teitelboim gravity  as being equivalent to general relativity (with an undetermined cosmological constant). This is true locally, but to be more precise, the unimodular condition imposes additional global constraints on spacetime.

In Henneaux--Teitelboim gravity a region of spacetime that is compact without boundary must have zero volume. Consequently, the metric must be degenerate everywhere. This is because
\begin{align}
    V_U = \int_U\td^4 x\,\sqrt{|g|} = \int_U\td^4 x\,\partial_{\mu}T^{\mu}= \int_{\partial U = \emptyset}\td^3x\,n_{\mu}T^{\mu} = 0\,.
\end{align}

One can compare this constraint to other formulations of unimodular gravity, where the metric determinant is fixed by $|g|\equiv 1$ or one chooses a fixed volume form $\omega$. In these cases the total volume of a compact spacetime is fixed to be $\int\td^4x \,1$ or $\int\omega$ respectively. In the same way, Henneaux--Teitelboim gravity fixes the volume of a compact spacetime without boundary to be zero.

In particular, we might consider a spacetime of the form $\Sigma\times S^1$, where $\Sigma$ is compact without boundary, as periodic in ``time''; such spacetimes must have a metric that is degenerate everywhere. As we already saw above, periodicity of a coordinate also affects the types of gauge fixing we can choose for $T^{\mu}$. If the $x^{\rho}$ coordinate parametrises a circle so that  we identify $x^\rho\sim x^\rho+L$, there is no gauge in which $T^{\rho}$ is the only non-zero component of $T^\mu$ unless the metric is again everywhere degenerate:
\begin{equation}
\int_0^L \td x^\rho\,\sqrt{|g|} = T^\rho|_{x^\rho=L}-T^\rho|_{x^\rho=0}=0\,.
\end{equation}

It is still possible to have a spacetime with non-degenerate metric that is periodic in some coordinate, as long as there is at least one non-compact direction. In that case, the total volume is infinite and unimodular time is not well-defined.

\subsection{Application: Cosmological Minisuperspace}
\label{minisupersec}

A simple example is that of a homogeneous isotropic Friedmann--Lema\^{i}tre--Robertson--Walker (FLRW) universe with metric given by
\begin{align}
\td s^2 = s\,N(t)^2\td t^2 + a(t)^2 \left[\td r^2 + S_k(r)^2\,\td\Omega^2\right]\,,
\end{align}
where $s=\pm 1$ indicates the choice of signature, $k\in\{+1,0,-1\}$ is a spatial curvature parameter,
\begin{align}
    S_k(r) = \begin{cases}\sin(r)\,, & k=1\,, \cr r\,, & k=0\,, \cr \sinh(r)\,, & k = -1\,,\end{cases}
\end{align}
and $\td\Omega^2$ is the standard line element on the 2-sphere. For this metric we have
\begin{align}
 R = 6\left(\frac{k}{a^2}-\frac{s\,\dot{a}^2}{ a^2N^2}+\frac{s\,\dot{a}\dot{N}}{aN^3}-\frac{s\,\ddot{a}}{ a N^2}\right)
\end{align}
and the vector density $T^\mu$ only depends on $t$ (and can be gauge-fixed to only have a component $T^0(t)$, see discussion above). After integrating over homogeneous constant $t$ hypersurfaces, the action (\ref{eq:HTaction}) becomes
\begin{align}
 S_{{\rm HT}}+S_{{\rm GHY}} & = \int_{t_0}^{t_1} \td t\left[3V_0\left(kNa-s\frac{\dot{a}^2 a}{ N}+s\frac{\dot{a}a^2\dot{N}}{ N^2}-s \frac{\ddot{a}a^2}{N}\right)-V_0 \Lambda N a^3+\Lambda\dot{T}\right]\nonumber
 \\ & \qquad +\left[3V_0\,s\frac{\dot{a} a^2}{ N}\right]_{t=t_0}^{t=t_1}\nonumber
  \\ & = \int_{t_0}^{t_1} \td t\left[3V_0\left(kNa+s\frac{\dot{a}^2 a}{ N}\right)-V_0 \Lambda N a^3+\Lambda\dot{T}\right]\,,
  \label{FLRWaction}
\end{align}
where $V_0$ is the volume of constant $t$ hypersurfaces using the ``fiducial metric'' 
\begin{align}
\td s^2_{{\rm fid}}=\td r^2 + S_k(r)^2\,\td\Omega^2
\end{align}
and we define $T=\int \td^3 x\;T^0$ as unimodular time in agreement with (\ref{Tidef}).
The resulting equations of motion are
\begin{align}
-s\frac{\dot{a}^2}{a^2 N^2}+\frac{k}{a^2}&=\frac{\Lambda}{3}\,,
\label{fried1}
\\ -s \frac{\ddot{a}}{a N^2}+s \frac{\dot{a}\dot{N}}{a N^3}& = \frac{\Lambda}{3}\,,
\label{fried2}
\\ \dot{T} & = V_0\,Na^3\,, 
\label{Tdotconst}
\\ \dot{\Lambda} & = 0\,.
\end{align}
The last equation tells us that $\Lambda$ is a constant, and the first two are then the regular Friedmann equations of cosmology. The $\dot{T}$ equation means that 
\begin{align}
T(t=t')-T(t=t'')=V_0\int_{t''}^{t'}\td t\,Na^3\,,
\end{align}
which again represents the spacetime volume between hypersurfaces $\Sigma(t=t'')$ and $\Sigma(t=t')$.

If we replace $\dot{T}$ in the action using (\ref{Tdotconst}), the last two terms vanish and $\Lambda$ disappears from the theory. Variation with respect to $N$ would then give us the first Friedmann equation with $\Lambda=0$ rather than $\Lambda$ as an undetermined constant. A more consistent reduction is obtained by viewing (\ref{Tdotconst}) as a possible gauge fixing condition for the reparametrisation invariance of a minisuperspace model, the type of partial fixing of diffeomorphism symmetry usually associated with ``unimodular'' gravity. 

If we posit that all variables are now parametrised by $T$ instead of the general coordinate $t$, we have $\dot{T}=1$ and hence $N=1/(V_0 a^3)$. This relation does not involve derivatives and can be substituted back into (\ref{FLRWaction}), yielding
\begin{align}
\label{eq:FLRWgf}
S_{{\rm gf}} = 3\int_{T_0}^{T_1} \td T\left[\frac{k}{a^2}+s V_0^2\dot{a}^2 a^4\right]
\end{align}
which is now an action for $a(T)$ alone; $N$ and $\Lambda$ no longer appear. The equation of motion is
\begin{align}
s\,\ddot{a}a^5+2s\,\dot{a}^2 a^4+\frac{1}{V_0^2}\frac{k}{a^2} = 0
\label{reducedEOM}
\end{align}
which is equivalent to the previous two Friedmann equations (\ref{fried1}) and (\ref{fried2}) in unimodular gauge. To see this, note that (\ref{reducedEOM}) can be multiplied by $\dot{a}/a$ and integrated to give
\begin{align}
 -s\dot{a}^2a^4+\frac{1}{V_0^2}\frac{k}{a^2}={\rm const.}
\end{align}
Up to a multiplicative factor, the constant is exactly the integration constant $\Lambda$ appearing in (\ref{fried1}). At least in this minisuperspace setting, we see that it is possible to formulate general relativity (with undetermined cosmological constant) as a fully gauge-fixed theory with no remaining coordinate freedom. One could see this as another particular feature of unimodular gravity.

For later comparison, it is instructive to write down the classical solutions for Euclidean signature and with closed spatial slices ($s=k=+1$). In cosmic time $N=1$ these are given by
\begin{equation}
    a(t)=\sqrt{\frac{3}{\Lambda}}\sin\left(\sqrt{\frac{\Lambda}{3}}t\right)
\end{equation}
for $\Lambda\neq 0$, and $a(t)=t$ for $\Lambda=0$. If $\Lambda>0$, there are two distinct solutions for given boundary values $a_1$ and $a_2$, which are either on the same or on different sides of the maximum $a_{{\rm max}}=\sqrt{\frac{3}{\Lambda}}$.

For $\Lambda>0$, the difference in unimodular time between the initial and final hypersurfaces evaluated on a classical solution is given by 
\begin{align}
    T(a=a_2)-T(a=a_1)&=2\pi^2\int_{t(a=a_1)}^{t(a=a_2)}\td t\;a^3(t)=\frac{3\pi^2}{2\Lambda^2}\left[\cos(\sqrt{3\Lambda}t)-9\cos\left(\sqrt{\frac{\Lambda}{3}}t\right)\right]_{t(a=a_1)}^{t(a=a_2)}\nonumber
    \\ & = \frac{2\pi^2}{\sqrt{3}\Lambda^2}\left((6+a_1^2\Lambda)\sqrt{3-a_1^2\Lambda}\pm(6+a_2^2\Lambda)\sqrt{3-a_2^2\Lambda}\right)
\label{eq:solapm}
\end{align}
where we take $a_2>a_1$ and the $\pm$ sign distinguishes between two independent solutions. One may verify that the same result holds for $\Lambda<0$, only with a single solution corresponding to the $-$ sign.

For $\Lambda=0$, there is again only a single solution interpolating between fixed boundary values and we find simply
\begin{align}
  T(a=a_2)-T(a=a_1)= 2\pi^2\int_{a_1}^{a_2}\td a\;a^3=\frac{\pi^2}{2}(a_2^4-a_1^4)\,.
\end{align}

\section{Unimodular Regge Calculus}

\begin{figure}
    \centering
    \includegraphics[width=0.5\linewidth]{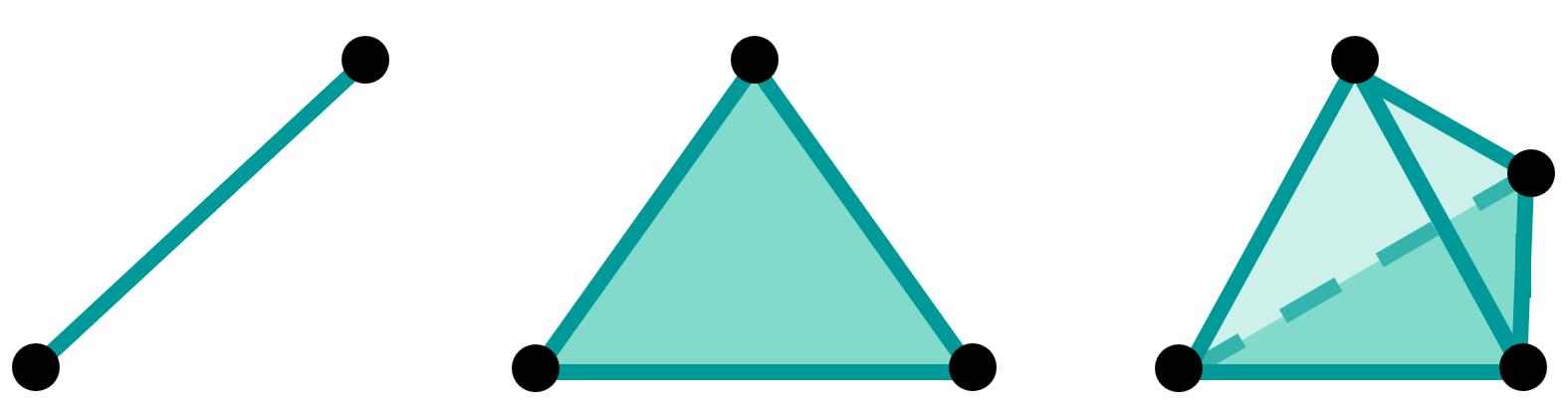}
    \caption{A $1$-dimensional simplex is two vertices connected by an edge. Adding a vertex and connecting it to the other two we get a triangle, i.e., a $2$-dimensional simplex. Adding another vertex we get a tetrahedron and if we would add one more we would get a $4$-simplex.}
    \label{fig:simplexvisual}
\end{figure}

Regge introduced the idea of a discrete theory of gravity \cite{Regge:1961px}. In Regge calculus, instead of  a smooth manifold where the curvature is continuous, one considers spacetime to be made up of discrete chunks, more precisely flat 4-simplices. A 4-simplex is the four-dimensional analogue of a triangle in two or a tetrahedron in three dimensions, see \cref{fig:simplexvisual}. It is defined by $\sigma = \{\sum_{i=1}^5t_iv_i|\sum_{i_1}^5t_i = 1\,,\,t_i>0\}$ where the $v_i$ are its five vertices. The boundary of a 4-simplex is formed by five tetrahedra, just as the boundary of a tetrahedron is formed by four triangles. The advantage of using this flat simplicial construction over using general higher-dimensional analogues of polyhedra is that the geometry is uniquely determined by the edge lengths. This is the crux of Regge's construction: instead of using coordinates and a metric, we can describe the discretised geometry by using edge lengths. 

\begin{figure}
    \centering
    \includegraphics[width=0.35\linewidth]{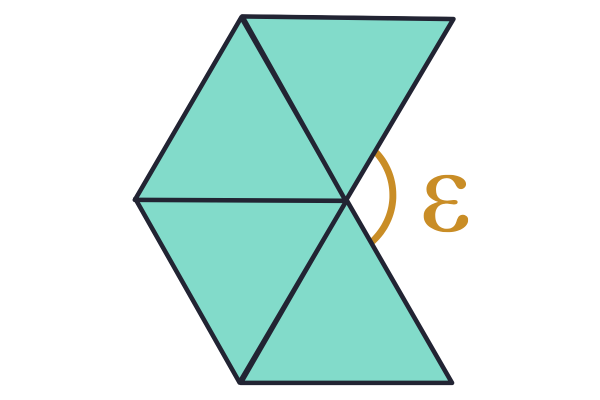}
    \caption{The deficit angle $\epsilon$ of a vertex $v$ is given by $2\pi$ minus the sum of the angles of the triangles at $v$.}
    \label{fig:deficitangle}
\end{figure}

If the metric can be replaced by the edge lengths, what is curvature in this discretised scenario? To understand the answer, let us for a moment consider a two-dimensional spacetime discretised using triangles. In flat space the angles in a triangle have to add up to $\pi$. However, if spacetime is positively or negatively curved, the angles should add up to more than $\pi$ and less than $\pi$ respectively. The triangles we use to discretise our spacetime are assumed to be flat, with angles adding up to $\pi$. Thus if we pick a vertex $v$, consider all triangles $t_i$ that have $v$ as a vertex, and add up the angles of these triangles at that vertex, the result will differ from $2\pi$ if there is curvature at that point. We call the difference between the sum and $2\pi$ the deficit angle $\epsilon$, see \cref{fig:deficitangle}. In two dimensions, curvature lives at vertices because a vertex can be placed inside a loop that measures  a deficit angle. The same idea generalises to $d$ dimensions: curvature is concentrated on $(d-2)$-simplices, called hinges or bones in Regge calculus, because these are the highest-dimensional objects around which you can form a closed loop in a plane orthogonal to them.


While subsequent advancements have built upon Regge’s work, see for instance \cite{Regge:2000wu, Barrett:2018ybl}, the core principles discussed in the following subsections are largely contained within the original paper.

\subsection{Fundamentals of Regge Calculus}
\label{sec:regge}

The majority of the literature on Regge calculus works in Euclidean signature. The reason for this is that using Lorentzian signature introduces many new challenges (e.g. defining Lorentzian angles \cite{Sorkin:2019llw}). While there has been some research in this direction \cite{Tate:2011ct,Dittrich:2021gww,Asante:2021phx,Ito:2022ycc,Borissova:2023izx}, we will stick to the Euclidean signature to simplify the calculations. 

The Regge action is defined on a given discrete structure built out of 4-simplices, tetrahedra, triangles, edges and vertices. Such a structure is called a simplicial complex and we will denote it by $\mathcal{T}$. A simplicial complex includes all simplices as well as the information of how they are glued together. We will denote the topological boundary of $\mathcal{T}$ by $\partial\mathcal{T}$. For example, the boundary of a 4-simplex is the union of its five boundary tetrahedra, and the boundary of two 4-simplices that share a tetrahedron is given by the union of eight tetrahedra. All triangles, edges and vertices that are fully contained within $\partial\mathcal{T}$ are defined to be boundary simplices. In the example of two 4-simplices, the shared tetrahedron is not part of the boundary, and we call it a bulk tetrahedron. Similarly every triangle, edge or vertex that is not a boundary simplex is called a bulk triangle, edge or vertex respectively. We will denote the bulk of the simplicial complex as $\mathring{\mathcal{T}}$, such that $\mathring{\mathcal{T}} = \mathcal{T}\backslash\partial\mathcal{T}$. The action of Regge calculus is given by
\begin{align}
    S[l_e] = \sum_{t\in\mathring{\mathcal{T}}}a_t(l_e)\epsilon_t(l_e)-\Lambda\sum_{\sigma}V_{\sigma}(l_e)+\sum_{t\in\partial\mathcal{T}}a_t(l_e)\epsilon^*_t(l_e)
    \label{eq:Reggeac}
\end{align}
where $a_t$ is the area of the triangle $t$, $V_{\sigma}$ is the 4-volume of the 4-simplex $\sigma$ and $l_e$ is the length of the edge $e$. The Regge action is a function of all edge lengths in the bulk and in the boundary. The deficit angle at a hinge $t$ is given by $\epsilon_t = 2\pi-\sum_{\sigma\supset t}\theta^{\sigma}_t$, where $\theta^{\sigma}_t$ is the dihedral angle in the simplex $\sigma$ located at the hinge $t$. Another way to view $\theta^{\sigma}_t$ is to project the simplex $\sigma$ to the plane orthogonal to $t$, so that $t$ is just a point $p_t$ and the two tetrahedra $\tau_{1/2}$ that are part of $\sigma$ and meet at $t$ are edges $e_{\tau_{1/2}}$. The angle $\theta^{\sigma}_t$ is then the angle between $e_{\tau_1}$ and $e_{\tau_2}$ in this plane. The last term in (\ref{eq:Reggeac}) is a boundary term analogous to the Gibbons--Hawking--York boundary term in the continuum \cite{Hartle:1981cf}, here the deficit angle is given by $\epsilon^*_t = \pi-\sum_{\sigma\supset t}\theta^{\sigma}_t$. 

The equations of motion are calculated from varying the action with respect to the edge lengths in the bulk only and are given by
\begin{align}
\label{eq:eomregge}
    \sum_{t\supset e}\frac{\partial a_t}{\partial l_e}\epsilon_t - \Lambda \sum_{\sigma\supset e}\frac{\partial V_{\sigma}}{\partial l_e} = 0\,.
\end{align}
The variation of the deficit angles vanishes due to the Schläfli identity $\sum_{t\subset \sigma}a_t\delta\theta^{\sigma}_t = 0$.

The usefulness of Regge calculus comes from its convergence to general relativity in the continuum limit, which has been shown in several ways. First it has been shown that solutions to the Euclidean Regge action converge to continuum solutions of general relativity \cite{Rocek:1981ama,Rocek:1982tj,Barrett:1988wd,Wong:1971,Gentle:1998,Brewin:2000zh}. Additionally, \cite{Cheeger:1983vq,Feinberg:1984he} proved the convergence of the Regge curvature to the Riemannian curvature. If a manifold considered in general relativity is a piecewise flat triangulation, the Einstein--Hilbert action becomes the Regge action \cite{Friedberg:1984ma}. This question has also been studied in the Lorentzian case \cite{Sorkin:1975}. 

Regge calculus is motivated by its applications to quantum gravity, where a discrete path integral may offer better convergence or mathematical behaviour than its continuous counterpart. Quantum Regge calculus was first explored in \cite{Rocek:1981ama,Rocek:1982tj,Hamber:1992jh,Hamber:1999nu,Loll:1998aj}. For some recent progress in this field see, e.g., \cite{Dittrich:2021gww,Asante:2021phx,Borissova:2023izx}. Possible quantum gravity applications are also our motivation for defining a Henneaux--Teitelboim version of Regge calculus. 

One particular challenge for a path integral formulation of Regge calculus is the existence of degenerate configurations with arbitrarily negative action, which may therefore dominate the path integral. Under a scaling of all edge lengths $l_e\mapsto Nl_e$ the dihedral angles stay invariant, so if the 4-volume is zero the action scales as $S[Nl_e] = N^2S[l_e]$, allowing it to become arbitrarily negative if the original action was negative. This is similar to the conformal factor problem in the continuous path integral and was already mentioned in \cite{Hartle:1985wr}. This issue is less problematic in approaches such as EDT or CDT, where the edge lengths are assumed to be fixed and thus at least these scaling divergences are avoided.

\subsection{Simplicial Discretisation of the Henneaux--Teitelboim Action}
\label{sec:uniregge}

In this section we want to make the Regge action unimodular or, put another way, discretise the Henneaux--Teitelboim action. First we have to allow $\Lambda$ to vary across different 4-simplices, such that for each $\sigma$ we have a value $\Lambda_{\sigma}$. Secondly we want to discretise the term including $T^{\mu}$. Let us assume that we have a triangulation $\mathcal{T}(M)$ of some manifold $M$, then
\begin{align}
    \int_{M}\td^4x\,\Lambda\partial_{\mu}T^{\mu} &= \sum_{\sigma\in \mathcal{T}(M)}\int_{\sigma}\td^4x\,\Lambda\partial_{\mu}T^{\mu}\nonumber\\
    &\approx \sum_{\sigma\in \mathcal{T}(M)}\Lambda_{\sigma}\int_{\sigma}\td^4x\,\partial_{\mu}T^{\mu}\nonumber \\
    &=  \sum_{\sigma\in \mathcal{T}(M)}\Lambda_{\sigma}\sum_{\tau\subset\sigma}\int_{\tau}\td^3x\,n_{\mu}T^{\mu}\,.
\end{align}
In the second step we have assumed that our triangulation is fine enough that $\Lambda$ can be assumed to be constant in every 4-simplex. The vector $n_{\mu}$ is normal to $\tau$ and outward pointing with respect to $\sigma$. This calculation motivates us to define $T^{\tau}_{\sigma} = \int_{\tau}\td^3x\,n_{\mu}T^{\mu}$. This gives us two new variables per bulk tetrahedron $\tau$, $T^{\tau}_{\sigma}$ and $T^{\tau}_{\sigma'}$, where $\sigma$ and $\sigma'$ are the two 4-simplices meeting at $\tau$. The only difference between the two variables is the sign of $n_{\mu}$, therefore $T^{\tau}_{\sigma} = -T^{\tau}_{\sigma'}$. We obtain one new variable $T^{\tau}_{\sigma}$ per boundary tetrahedron.

Putting all this together we can write down the new {\em Henneaux--Teitelboim--Regge} action
\begin{align}
\begin{split}
\label{eq:umreggeaction}
     S_{{\rm HTR}}[l_e,\Lambda_{\sigma},T^{\tau}_{\sigma},\mu_{\tau}] =& \sum_{t\in\mathring{\mathcal{T}}}a_{t}(l_e)\epsilon_{t}(l_e)-\sum_{\sigma}\Lambda_{\sigma}\left[V_{\sigma}(l_e)-\sum_{\tau\subset\sigma}T^{\tau}_{\sigma}\right]\\&+\sum_{\tau\in\mathring{\mathcal{T}}}\mu_{\tau}\sum_{\sigma\supset\tau}T^{\tau}_{\sigma}+\sum_{t\in\partial\mathcal{T}}a_t(l_e)\epsilon^*_t(l_e)\,.
\end{split}
\end{align}
Fixed boundary data for the action are given by $\{l_e,T^{\tau}_{\sigma}\}_{e,\tau \in \partial \mathcal{T}}$. Apart from the variation with respect to the bulk edge lengths $l_e$ we can now also vary with respect to the bulk variables $\Lambda_{\sigma}$, $T^{\tau}_{\sigma}$ or $\mu_{\tau}$. The latter variation leads to the following condition for bulk tetrahedra:
\begin{align}
\label{eq:tcon}
    \sum_{\sigma\supset\tau}T^{\tau}_{\sigma} = 0\,,
\end{align}
which ensures that the variables $T^\tau_\sigma$ have a consistent interpretation as discrete analogues of $\int_{\tau}\td^3x\,n_{\mu}T^{\mu}$.
Varying $T^{\tau}_{\sigma}$ for $\tau\in\mathring{\mathcal{T}}$ gives us the equation
\begin{align}
\label{eq:lambdacon}
    \Lambda_{\sigma}+\mu_{\tau} = 0\,,\quad\tau\subset\sigma\,.
\end{align}
This constrains $\Lambda_{\sigma} = \Lambda_{\sigma'}$ if $\sigma$ and $\sigma'$ share a tetrahedron. By connectivity this means that $\Lambda_{\sigma} = \Lambda$ is the same in every 4-simplex within the same connected component (and we may assume the overall complex is connected). The unimodular condition comes from varying with respect to $\Lambda_{\sigma}$:
\begin{align}
\label{eq:unicondis}
    V_{\sigma}(l_e)-\sum_{\tau\subset\sigma}T^{\tau}_{\sigma} = 0\,.
\end{align}
Just as in the continuum version, we have a gauge freedom in $T^{\tau}_{\sigma}$: we can transform  $T^{\tau}_\sigma\rightarrow T^\tau_\sigma + \xi^{\tau}_{\sigma}$ under the condition that (\ref{eq:unicondis}) and (\ref{eq:tcon}) still hold. The exact number of degrees of freedom in the system depends on the particular simplicial complex we are considering. As an example, for a simplicial complex obtained from discretising a continuum manifold, any $\xi^\mu$ with $\partial_\mu \xi^\mu=0$ can be used to define $\xi^{\tau}_{\sigma} = \int_{\tau}\td^3x\,n_{\mu}\xi^{\mu}$ that satisfy
\begin{equation}
\sum_{\tau\subset\sigma}\xi^{\tau}_{\sigma} = \sum_{\tau\subset\sigma} \int_{\tau}\td^3x\,n_{\mu}\xi^{\mu} = \int_{\sigma}\td^4x\,\partial_{\mu}\xi^{\mu} = 0\,,
\end{equation}
hence leaving the constraint equations unchanged.

Variation with respect to the bulk edge lengths gives
\begin{align}
    \sum_{t\supset e}\frac{\partial a_t}{\partial l_e}\epsilon_t - \sum_{\sigma\supset e}\Lambda _\sigma\frac{\partial V_{\sigma}}{\partial l_e} = 0\,.
\end{align}
Using the constraint (\ref{eq:lambdacon}) which implies that $\Lambda_\sigma$ is constant, these are the standard Regge equations of motion.

All this is analogous to Henneaux--Teitelboim gravity in the continuum. We also see that the volume of a simplicial complex $\mathcal{T}_0$ without a boundary is zero:
\begin{align}
    V_{\mathcal{T}_0} = \sum_{\sigma\in\mathcal{T}_0}V_{\sigma} = \sum_{\sigma\in\mathcal{T}_0}\sum_{\tau\subset\sigma}T^{\tau}_{\sigma} = 0\,.
    \label{eq:zerovolume}
\end{align}
This is because for every $\tau\in\mathring{\mathcal{T}_0}$ both $T^{\tau}_{\sigma}$ and $T^{\tau}_{\sigma'}$ appear in the sum, where $\sigma$ and $\sigma'$ are the two 4-simplices that share $\tau$. And since $\mathcal{T}_0$ has no boundary, every tetrahedron is a bulk tetrahedron, and everything cancels out. Since the total volume is additive, this means that every 4-simplex then has a vanishing volume, which may be realised by different kinds of degenerate configurations in terms of edge lengths.

We can also define a discrete analogue of unimodular time. Given a simplicial complex $\mathcal{T}$ with boundary $\partial\mathcal{T} = \Sigma_1\sqcup\Sigma_2$, we may call the $\Sigma_i$'s hypersurfaces and declare $\Sigma_1$ to be the initial hypersurface and $\Sigma_2$ to be the final hypersurface. This choice fixes the arrow of ``time", again with the understanding that in Euclidean signature this just means a general evolution parameter. (Even in the Lorentzian case, we can generalise ``time'' to mean a general evolution parameter, as discussed in \cref{sec:continuum}.) In analogy with (\ref{Tidef}) we then define the unimodular time on $\Sigma_1$ and $\Sigma_2$ as
\begin{align}
    T_i = (-1)^i\sum_{\tau\in\Sigma_i}T^{\tau}_{\sigma}\,.
    \label{discrete_Tidef}
\end{align}
The sign comes from the interpretation of $T^{\tau}_{\sigma}$ as a discrete analogue of $\int_{\tau}\td^3x\,n_{\mu}T^{\mu}$. The outward pointing normal $n_{\mu}$ is defined to be future-directed if $\sigma$ is in the past of $\tau$. This is the case for $\tau\in\Sigma_2$. However, for $\tau\in\Sigma_1$, we would interpret $n_\mu$ as pointing in the past direction, and we need to invert it by multiplying with $-1$.

This definition only refers to parts of the boundary of a simplicial complex, whereas in the continuum unimodular time could be defined for arbitrary hypersurfaces. However, consider two separate simplicial complexes of the form we just discussed, where $\mathcal{T}$ has a boundary consisting of the initial hypersurface $\Sigma_1$ and final hypersurface $\Sigma_2$, and $\mathcal{T}'$ has a boundary consisting of the initial hypersurface $\Sigma'_2$ and final hypersurface $\Sigma_3$. $\Sigma_2$ and $\Sigma'_2$ are assumed to be combinatorially identical, have the same edge lengths and variables $T^\tau$ that are of opposite sign, i.e.,
\begin{equation}
T^\tau_\sigma = -T^{\tau'}_{\sigma'}
\label{consistencysign}
\end{equation}
if $\tau\in\Sigma_2$ and $\tau'\in\Sigma'_2$ correspond to the ``same'' tetrahedron in the two hypersurfaces, and $\tau\subset\sigma\in\mathcal{T}$ while $\tau'\subset\sigma'\in\mathcal{T}'$. In this case, $\Sigma_2$ and $\Sigma'_2$ have the same value of unimodular time.

In such a case, the two simplicial complexes can be joined along $\Sigma_2$ and $\Sigma_2'$ to form a new simplex $\mathcal{T}''$ with initial hypersurface $\Sigma_1$ and final hypersurface $\Sigma_3$: identical boundary tetrahedra $\tau$ and $\tau'$ in $\Sigma_2$ and $\Sigma_2'$ become identified as a single new bulk tetrahedron shared by 4-simplices $\sigma$ and $\sigma'$, with $T^\tau_\sigma$ and $T^{\tau'}_{\sigma'}$ associated to this bulk tetrahedron, and (\ref{consistencysign}) guarantees consistency. The new ``hypersurface'' formed by all newly defined bulk tetrahedra can also be assigned with a value of unimodular time, equal to the one previously assigned to $\Sigma_2$ and $\Sigma'_2$. Hence, ``gluing'' simplical complexes allows defining unimodular time also for tetrahedra that are not in the boundary of the final simplicial complex. This is particularly relevant if we want to add more and more intermediate time steps and explore a continuum limit in which this notion of unimodular time can converge to the one of the continuum.

The definition of unimodular time is independent of the discrete structure in between the hypersurfaces. The simplest possibility is that the four-dimensional building blocks are simply the three-dimensional ones times an interval $\sigma =\tau\times I$ (allowing for a generalised construction in which the $\sigma$ do not form a simplicial structure), but one may also imagine more complicated structures between hypersurfaces. An example of both $\sigma = \tau \times I$ and a true simplicial structure will be shown in the next section.

Just like in the continuum, unimodular time has an interpretation as a volume time. Let us assume $\Sigma_i$ to be compact without boundary. The volume of $\mathcal{T}$ is then given by
\begin{align}
    V_{\mathcal{T}} = \sum_{\sigma\in \mathcal{T}}V_{\sigma} = \sum_{\sigma\in \mathcal{T}}\sum_{\tau\subset\sigma}T^{\tau}_{\sigma} = \sum_{\tau\in\partial\mathcal{T}}T^{\tau}_{\sigma} = T_{2}-T_1\,.
\end{align}

We saw in (\ref{eq:zerovolume}) that, just as in the continuum formulation of Henneaux--Teitelboim gravity, the only solutions for compact spacetimes without boundary are geometries that are degenerate everywhere. Such configurations will be subject to the problem of large negative action that we described in \cref{sec:regge}, so extra care is needed if such topologies are meant to be included in a proposed path integral definition. In EDT and CDT this pathology is dealt with by fixing the edge lengths. However, the usual formulation of EDT takes not only the edge lengths to be fixed, but also the topology of spacetime, typically taking it to be $S^4$ \cite{Ambjorn:2013eha,Asaduzzaman:2022kxz}. This seems incompatible with the Henneaux--Teitelboim--Regge formulation, where all solutions with topology $S^4$ are degenerate. Historically there have been attempts to study 4-dimensional manifolds with boundary in EDT, but the theory failed to produce an extended 4-dimensional spacetime \cite{Ambjorn:2005jj}. In CDT additional (causality) conditions are imposed on the simplicial spacetimes in the path integral, making the study of transition amplitudes between 3-geometries possible. In particular CDT imposes a foliation of the discrete spacetime, which gives an inbuilt geometric notion of proper time \cite{Loll:2019rdj}. 

A major difference between unimodular time and a notion of time coming from the geometry of the 4-dimensional building blocks is that unimodular time is a variable on the boundaries. Thus one can impose a boundary condition of how much time elapses between two hypersurfaces. In a quantum theory where boundary observables are represented as operators on a boundary Hilbert space, one could potentially study the spectrum of the unimodular time operator. Because unimodular time is not a coordinate but a field itself, it allows for the introduction of time into the quantum theory in a way that is not possible in quantisations of the Einstein-Hilbert action alone.

\section{Applications and Models}
\label{sec:models}

\begin{figure}
    \centering
    \includegraphics[width=0.5\linewidth]{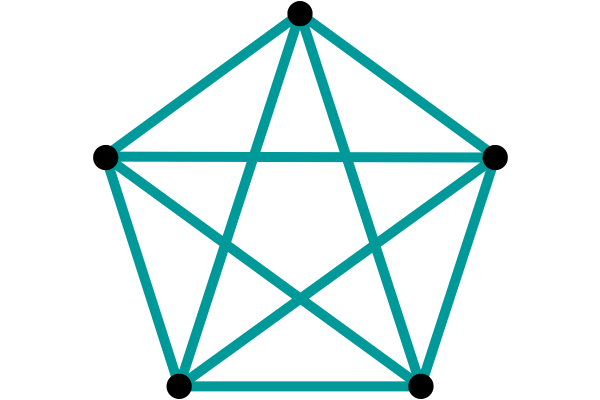}
    \caption{A 4-simplex has 5 vertices and maximal connectivity. It has 5 boundary tetrahedra, 10 edges and 10 triangles. The boundary of a 4-simplex is a regular triangulation of a 3-sphere, called the 5-cell. The boundary graph is its own dual, meaning if we draw a vertex for each tetrahedron and connect the vertices whose tetrahedra share a triangle we again have a 5-cell. Thus one may also visualise the boundary by taking every vertex in the image above to stand for a tetrahedron and the edges describing how these are glued together. }
    \label{fig:4simplex}
\end{figure}

To better understand how the modifications to the Regge action that we have made affect calculations, we will have a closer look at two example spacetimes. Both example complexes $\mathcal{T}_{cos}$ and $\mathcal{T}_{sim}$ have the same boundary $\partial\mathcal{T}_{cos} = \partial\mathcal{T}_{sim} = \partial\Delta^4\sqcup\partial\Delta^4$, where $\Delta^4$ denotes the standard 4-simplex. Its boundary $\partial\Delta^4$ triangulates the 3-sphere with five tetrahedra and is called the 5-cell. A visualisation of this triangulation is shown in \cref{fig:4simplex}.
We will further assume in both examples that all edge lengths in one of the boundary hypersurfaces $\partial\Delta^4$ are the same, so that each hypersurface forms a regular triangulation of the 3-sphere. The difference between the models is in the bulk of the complex. In our first example, which we will call the cosmological model, we match each vertex in the initial 5-cell to one in the final 5-cell and connect them. The 4-dimensional building blocks are thus simply $\sigma = \tau\times I$, which is a four-dimensional frustum instead of a 4-simplex. This model has been studied in both Euclidean\footnote{In this paper the authors use Euclidean angles, which is why we classify it as a Euclidean treatment. However, they use imaginary time which leads to their equations being slightly different from ours.} \cite{Liu:2015gpa} and Lorentzian \cite{Dittrich:2021gww} signature Regge calculus. We will compare our results to the literature in the next section.

On the other hand, our second model, which we call the simplicial model,  is truly a simplicial complex, as the 4-dimensional building blocks are 4-simplices. We construct the simplicial complex by again choosing for each vertex in the initial 5-cell a vertex in the final 5-cell and then connecting the initial vertex to every vertex in the final hypersurface except the chosen one. 

We will first discuss the cosmological model.

\subsection{Discrete Cosmological Model}
\label{sec:cosmomodel}
\begin{figure}
    \centering
    \includegraphics[width=0.75\linewidth]{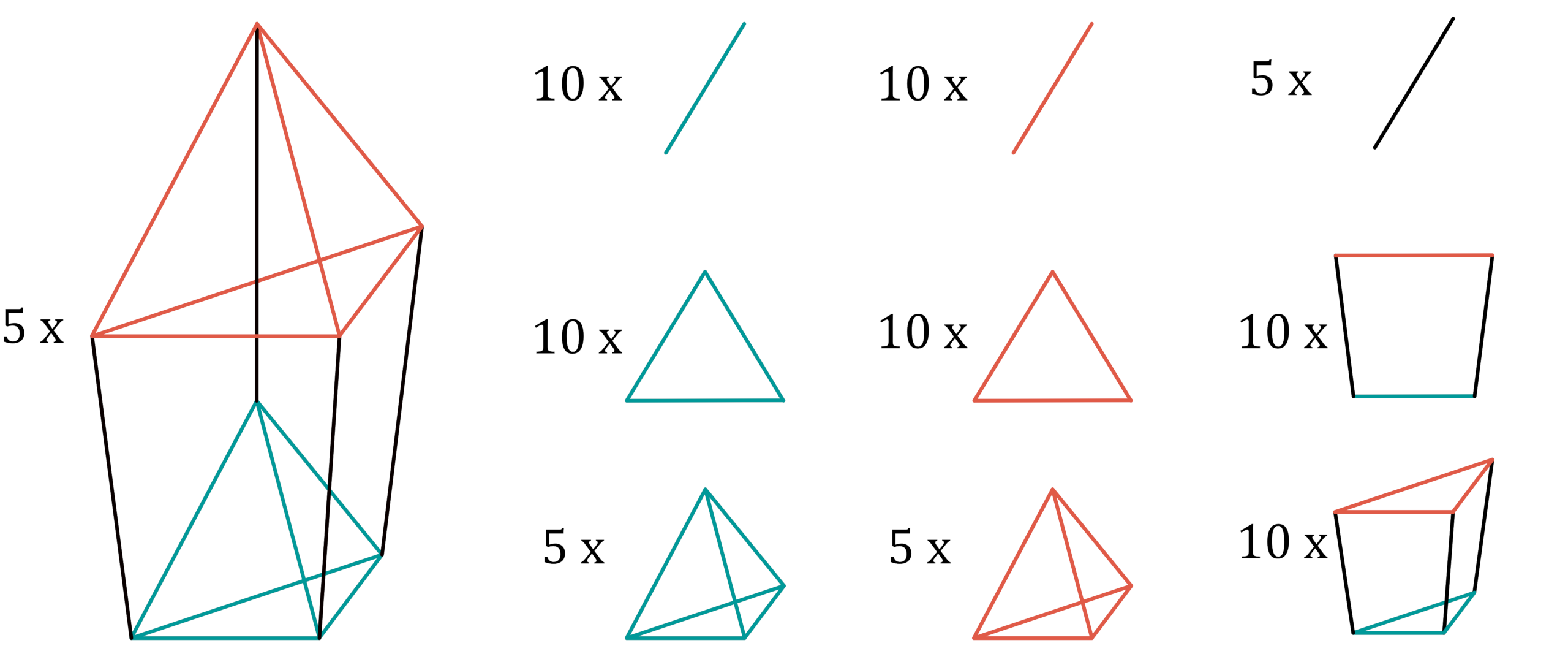}
    \caption{$\mathcal{T}_{cos}$ is made up of five 4-dimensional frusta, visualised on the left. On the right-hand side we list the lower dimensional building blocks (cells, faces and edges), indicating the multiplicity of each component within the overall structure.}
    \label{fig:cosmobb}
\end{figure}

We denote the vertices of the initial 5-cell by $\{x_i\}_{1\leq i\leq 5}$ and the final vertices by $\{y_i\}_{1\leq i\leq 5}$. As stated above, we take all edges connecting $x$ vertices to have the same length, which we will call $l_1$. All edges connecting $y$ vertices are taken to have length $l_2$. To connect the two hypersurfaces, we connect the vertex $x_i$ to the vertex $y_i$. We also assume all edges connecting $x$ and $y$ vertices to have the same length, which we denote by $m$. A simplex is then fully characterised by the number of $x$ and $y$ vertices it contains. We write, e.g., $t_{22}$ for a bulk trapezium with two $x$ and two $y$ vertices. A visualisation of building blocks of the complex is shown in \cref{fig:cosmobb}.

To write down the Regge action (now generalised beyond its initial simplicial definition), we need to calculate the areas of all faces, the 4-volume of all frusta and the dihedral angles associated to all hinges. Let us start with the areas of the faces:
\begin{align}
    a_{30} = \frac{\sqrt{3}}{4}l_1^2\,,\quad a_{03} = \frac{\sqrt{3}}{4}l_2^2\,,\quad a_{22} = \frac{l_1+l_2}{2}\sqrt{m^2-\frac{1}{4}(l_2-l_1)^2} = \frac{l_1+l_2}{2}\sqrt{h^2+\frac{1}{8}(l_2-l_1)^2}\,.
\end{align}
In terms of the building blocks illustrated in \cref{fig:cosmobb}, these areas correspond to the areas of a boundary triangle in the initial hypersurface, a boundary triangle in the final hypersurfaces and a bulk trapezium. We have written the formula for $a_{22}$ both in terms of the edge length $m$ and the height of the 4-dimensional building block $h = \sqrt{m^2-\frac{3}{8}(l_2-l_1)^2}$. The volume of a frustum $\sigma_{44}$ is given by
\begin{align}
    V_{44} = \frac{h}{24\sqrt{2}}(l_1^3+l_1^2l_2+l_1l_2^2+l_2^3)\,.
\end{align}
The dihedral angles at a hinge $t$ can be calculated, e.g., by embedding the 4-dimensional building block into $\mathbb{R}^4$ and then projecting everything onto the surface normal to $t$. The dihedral angles between tetrahedra are thereby turned into two-dimensional angles which can be calculated as usual. Following this recipe, we find the following dihedral angles:
\begin{align}
    \theta_{30}  &= \cos^{-1}\left(\frac{l_1-l_2}{2\sqrt{2}\sqrt{3m^2-(l_2-l_1)^2}}\right) = \cos^{-1}\left(\frac{l_1-l_2}{\sqrt{24h^2+(l_2-l_1)^2}}\right)\,,\\
     \theta_{03} &= \cos^{-1}\left(\frac{l_2-l_1}{2\sqrt{2}\sqrt{3m^2-(l_2-l_1)^2}}\right) = \cos^{-1}\left(\frac{l_2-l_1}{\sqrt{24h^2+(l_2-l_1)^2}}\right)\,,\\
    \theta_{22} &= \sec^{-1}\left(\frac{2m^2}{2m^2-(l_2-l_1)^2}+2\right) = \cos^{-1}\left(\frac{8h^2-(l_2-l_1)^2}{24h^2+(l_2-l_1)^2}\right)\,.
\end{align}
The expressions for the areas, the 4-volume and the angle $\theta_{22}$ are the same as those stated in \cite{Liu:2015gpa,Dittrich:2021gww}. The expressions for $\theta_{30}$ and $\theta_{03}$ are slightly different from the ones in \cite{Liu:2015gpa}, where the authors use imaginary time leading to some sign differences. In Lorentzian Regge calculus, the dihedral angles for $t_{30}$ and $t_{03}$ would be very different from their Euclidean counterparts \cite{Sorkin:2019llw}. To calculate the deficit angles, we need to consider the combinatorics of how many four-dimensional building blocks a hinge (triangle or trapezium) is part of. We end up with
\begin{align}
    \epsilon^*_{30} &= 2\sin^{-1}\left(\frac{l_1-l_2}{2\sqrt{2}\sqrt{3m^2-(l_2-l_1)^2}}\right) = 2\sin^{-1}\left(\frac{l_1-l_2}{\sqrt{24h^2+(l_2-l_1)^2}}\right)\,,\\
    \epsilon^*_{03} &= 2\sin^{-1}\left(\frac{l_2-l_1}{2\sqrt{2}\sqrt{3m^2-(l_2-l_1)^2}}\right) = 2\sin^{-1}\left(\frac{l_2-l_1}{\sqrt{24h^2+(l_2-l_1)^2}}\right)\,,\\
    \epsilon_{22} &= 2\pi-3\sec^{-1}\left(\frac{2m^2}{2m^2-(l_2-l_1)^2}+2\right) = 2\pi-3\cos^{-1}\left(\frac{8h^2-(l_2-l_1)^2}{24h^2+(l_2-l_1)^2}\right)\,.
\end{align}
With all this, we can now write down the Henneaux--Teitelboim--Regge action for our cosmological model:
\begin{align}
\begin{split}
   S =&\, \frac{5(l_1+l_2)}{2}\sqrt{4m^2-(l_2-l_1)^2}\left(2\pi-3\sec^{-1}\left(\frac{2m^2}{2m^2-(l_2-l_1)^2}+2\right)\right)\\
    &+\frac{5\sqrt{3}}{2}\left[2l_1^2\sin^{-1}\left(\frac{l_1-l_2}{2\sqrt{2}\sqrt{3m^2-(l_2-l_1)^2}}\right)+2l_2^2\sin^{-1}\left(\frac{l_2-l_1}{2\sqrt{2}\sqrt{3m^2-(l_2-l_1)^2}}\right)\right]\\
     &-\frac{(l_1^2+l_2^2)(l_1+l_2)}{24\sqrt{2}}\sqrt{m^2-\frac{3}{8}(l_2-l_1)^2}\sum_{\sigma}\Lambda_{\sigma}+\sum_{\sigma}\Lambda_{\sigma}\sum_{\tau\subset\sigma}T^{\tau}_{\sigma}+\sum_{\tau_{33}}\mu_{\tau_{33}}\sum_{\tau_{33}\subset\sigma}T^{\tau_{33}}_{\sigma}
\end{split}\\
\begin{split}
    =&\,\frac{5}{2\sqrt{2}}(l_1+l_2)\sqrt{8h^2+(l_2-l_1)^2}\left(2\pi-3\sec^{-1}\left(\frac{4(l_2-l_1)^2}{8h^2-(l_2-l_1)^2}+3\right)\right)\\
    &+ \frac{5\sqrt{3}}{2}\left[2l_1^2\sin^{-1}\left(\frac{l_1-l_2}{\sqrt{24h^2+(l_2-l_1)^2}}\right)+2l_2^2\sin^{-1}\left(\frac{l_2-l_1}{\sqrt{24h^2+(l_2-l_1)^2}}\right)\right]\\
    &-\frac{h(l_1^2+l_2^2)(l_1+l_2)}{24\sqrt{2}}\sum_{\sigma}\Lambda_{\sigma}+\sum_{\sigma}\Lambda_{\sigma}\sum_{\tau\subset\sigma}T^{\tau}_{\sigma}+\sum_{\tau_{33}}\mu_{\tau_{33}}\sum_{\tau_{33}\subset\sigma}T^{\tau_{33}}_{\sigma}\,.
    \label{long_action}
\end{split}
\end{align}
We have written everything in terms of both the edge lengths $m$ and the height of the frusta $h$ for an easier comparison with  \cite{Liu:2015gpa,Dittrich:2021gww}. We can simplify the last line by solving the constraint (\ref{eq:lambdacon}) demanding $\Lambda$ to be a global constant. Further, the definition (\ref{discrete_Tidef}) means that the unimodular times of the initial and final hypersurfaces $\Sigma_1$ and $\Sigma_2$ are given by $T_1 = -\sum_{\tau\in\Sigma_1}T^{\tau}_{\sigma}$ and $T_2 = \sum_{\tau\in\Sigma_2}T^{\tau}_{\sigma}$. The last line of (\ref{long_action}) then becomes
\begin{align}
    -\frac{5h(l_1^2+l_2^2)(l_1+l_2)}{24\sqrt{2}}\Lambda +\Lambda (T_2-T_1)\,.
\end{align}
An interesting question is how our discrete model connects to the continuum theory, and how unimodular time may help perform this limit. Following the method of \cite{Dittrich:2021gww}, we  take the continuum time limit by making the following substitutions:
\begin{align}
\label{eq:tcont1}
\begin{split}
    l_1&\rightarrow l\,, \quad l_2 \rightarrow l+\dot{l}\,\td t\,,\\
    T_1 &\rightarrow T\,,\quad T_2\rightarrow T+\dot{T}\,\td t\,,\\
    h&\rightarrow N\td t\,.
\end{split}
\end{align}
Spatially, we do not take the limit of a finer triangulation of the 3-sphere, but we remain in a discrete model and relate the edge length $l$ to a continuum scale factor $a$. There is an ambiguity in how this identification works, and two possible ways were used in \cite{Dittrich:2021gww}. One can either calculate the 3-volume of the triangulation with edge length $l$, $V_{\mathcal{T}} = \frac{5l^3}{6\sqrt{2}}$, and fix the corresponding scale factor $a_V$ by requiring that this volume be equal to that of a continuum 3-sphere with radius $a_V$, $V_{S^3} = 2\pi^2a_V^3$. Another way is to calculate the three-dimensional integrated Ricci curvature of the 5-cell and require that it should equal the integrated curvature of a 3-sphere with radius $a_R$. The three-dimensional integrated curvature of a triangulation is given by $R_{\mathcal{T}} = 2\sum_e l_e\epsilon_e$, which for a regular 5-cell is given by $R_{\mathcal{T}} = 20l(2\pi-3\cos^{-1}(1/3))$. The integrated curvature of a 3-sphere with radius $a_R$ is given by $R_{S^3} = 12\pi^2a_R$. In both cases, a continuum 3-sphere is compared with its triangulated analogue, and so discretisation artifacts are carried through in the translation. Explicitly, for the 5-cell we have
\begin{align}
\label{eq:ltoa}
    l = 2^{\frac{5}{6}}\left(\frac{3\pi^2}{5}\right)^{\frac{1}{3}}a_V \approx 3.224\, a_V\,,\quad l = \frac{3\pi^2}{10\pi-15\cos^{-1}(1/3)}a_R \approx 2.286\, a_R\,.
\end{align}
Substituting (\ref{eq:tcont1}) into the action (\ref{long_action}), we find the continuous time Lagrangian
\begin{align}
    L(l,\dot{l},T,\dot{T},N,\Lambda) =& \lim_{\td t\rightarrow 0}\frac{S(l,l+\dot{l}\,\td t,T,T+\dot{T}\,\td t, N\,\td t, \Lambda)}{\td t} \\
    \begin{split}
    =& -\frac{5l}{\sqrt{2}}\sqrt{\dot{l}^2+8N^2}\left(\pi-3\cos^{-1}\left(\frac{\dot{l}^2-8N^2}{\dot{l}^2+24N^2}\right)\right)\\
    &-5\sqrt{3}l\dot{l}\left(\pi-2\cos^{-1}\left(\frac{\dot{l}}{\sqrt{\dot{l}^2+24N^2}}\right)\right)-\frac{5N}{6\sqrt{2}}l^3\Lambda+\Lambda\dot{T}\,.
    \end{split}
\end{align}
This expression does not resemble the continuum minisuperspace Lagrangians discussed in section \ref{minisupersec}, but it contains effectively infinitely high powers of the velocity $\dot{l}$. To get a meaningful comparison with (\ref{FLRWaction}) we can expand in orders of $\dot{l}/N$, as was done in \cite{Dittrich:2021gww}. This gives
\begin{align}
\label{eq:lag5cell}
     L(l,\dot{l},T,\dot{T},N,\Lambda) = c_1 lN + c_2 l\frac{\dot{l}^2}{N}-\frac{5N}{6\sqrt{2}}\Lambda l^3+\Lambda\dot{T} +\mathcal{O}\left(\left(\frac{\dot{l}}{N}\right)^3\right)\,,\\
     c_1 = 10\left(2\pi-3\cos^{-1}\left(\frac{1}{3}\right)\right)\,,\quad c_2 = \frac{5}{8}\left(2\sqrt{2}+2\pi-3\cos^{-1}\left(\frac{1}{3}\right)\right)\,.
\end{align}
This Lagrangian agrees with the one derived in \cite{Dittrich:2021gww}, apart from the sign in front of the $\dot{l}^2$ term and our new term $\Lambda\dot{T}$. These are the expected differences, given that we work in Euclidean and not in Lorentzian signature, and we have added an Henneaux--Teitelboim term to the action. To compare with the continuum Lagrangian (\ref{FLRWaction}), we now use either of the substitutions in (\ref{eq:ltoa}). This gives
\begin{align}
    L(a,\dot{a},T,\dot{T},N,\Lambda) = 6\pi^2\left[\tilde{c}_1 Na + \tilde{c}_2 \frac{a\dot{a}^2}{N}-\tilde{c}_3\Lambda N a^3+\tilde{c}_4\Lambda\dot{T} +\mathcal{O}\left(\left(\frac{\dot{a}}{N}\right)^3\right)\right]\,,
\end{align}
where the numerical factors are summarised in \cref{tab:compcont}. Our 5-cell model is not a very good approximation, as the difference between the continuum factors and ours is of the same order as the factors. However, we can see that using the scale factor $a_V$ we get slightly better agreement compared to using $a_R$. This may be because the notion of volume in the triangulation is similar to the notion of volume in the continuum, whereas the notion of curvature is radically changed and therefore more sensitive to the discrete structure. In table 3 of \cite{Dittrich:2021gww} the authors show that the more refined 600-cell model results in much better agreement with the continuum, and we would expect the same result here.
\begin{table}[]
    \centering
    \begin{tabular}{c|c|c|c|c}
         & $\tilde{c}_1$ & $\tilde{c}_2$ & $\tilde{c}_3$ & $\tilde{c}_4$ \\\hline
         $a_V$ & $1.410$ & $1.916$ & $\frac{1}{3}$ & $\frac{1}{6\pi^2}$ \\
         $a_R$ & $1$ & $0.683$ & $0.119$ & $\frac{1}{6\pi^2}$\\
         $a_{\text{cont.}}$ & $1$ & $1$ & $\frac{1}{3}$ & $\frac{1}{6\pi^2}$ 
    \end{tabular}
    \caption{Comparison of our Lagrangian in the continuous time limit to the continuum FLRW model in (\ref{FLRWaction}). Our factors $\tilde{c}_1$, $\tilde{c}_2$ and $\tilde{c}_3$ agree with the ones found in \cite{Dittrich:2021gww}.}
    \label{tab:compcont}
\end{table}

From (\ref{eq:lag5cell}) we can also calculate the equations of motion. Variation with respect to the lapse $N$ and edge length $l$ gives
\begin{align}
    \frac{5\Lambda}{6\sqrt{2}}l^3-c_1 l +c_2\frac{l\dot{l}^2}{N^2} = 0\,,\\
    c_2\frac{\dot{l}^2}{N}+2c_2l\,\frac{\td}{\td t}\left(\frac{\dot{l}}{N}\right)+\frac{5N}{2\sqrt{2}}\Lambda l^2-c_1 N = 0\,.
\end{align}
The first equation agrees with the Hamiltonian constraint derived in \cite{Liu:2015gpa} expanded to second order in $\dot{l}$, up to the sign in front of the $\dot{l}^2$ term and after setting $N=1$. The second equation is the acceleration equation and is consistent with the Hamiltonian constraint as they are derived from a Lagrangian. In \cite{Liu:2015gpa}, the authors show that their acceleration equation is consistent with their Hamiltonian constraint including all orders of $\dot{l}$. Thus our results are compatible with those of \cite{Liu:2015gpa}.

So far we have been able to reproduce known results in terms of the continuum limit of this cosmological model while including unimodular time. The use of unimodular time did not simplify the derivations, because we used the height $h$ of the frusta to define time evolution by taking $h\rightarrow N\td t$, and there is still an arbitrary lapse $N$. Alternatively, we can solve the unimodular constraint before taking the continuous time limit, and take $T$ to define temporal evolution. This was precisely the philosophy behind deriving (\ref{eq:FLRWgf}) in the continuum.

In our formulation we do not have a single unimodular constraint, but one for each four-dimensional frustum, see (\ref{eq:unicondis}). However, a lot of the free variables are manifestations of the gauge freedom in the continuum. If we instead look at the total 4-volume we have
\begin{align}
    V_{\mathcal{T}_{cos}} = \sum_{\sigma\in\mathcal{T}_{cos}}V_{\sigma} = \sum_{\sigma\in\mathcal{T}_{cos}}\sum_{\tau\subset\sigma}T^{\tau}_{\sigma} =  T_2 - T_1 = \Delta T\,.
\end{align}
As in previous calculations, the contributions from bulk frustum variables $T^{33}_{44}$ have cancelled in the sum, and we only have two degrees of freedom left that can be seen as boundary data. Calculating the total volume of the complex we can rewrite this constraint as
\begin{align}
    \Delta T = \frac{5h}{24\sqrt{2}}(l_1^3+l_1^2l_2+l_1l_2^2+l_2^3)\,.
    \label{deltaT}
\end{align}
Connecting to our discussion in the continuum, we see that we may only identify the initial and final hypersurface (and thus have a closed timelike loop) if $l_1 = l_2 = 0$. We now want to define a prescription for taking the continuous time limit in unimodular time. We replace
\begin{align}
    \Delta T \rightarrow \td T\,,\quad l_1\rightarrow l\,,\quad l_2\rightarrow l+\dot{l}\,\td T\,.
    \label{continuum_unimod}
\end{align}
The dot now symbolises the derivative with respect to unimodular time $T$. Rearranging (\ref{deltaT}) with (\ref{continuum_unimod}) gives us the following for $h$ at leading order in $\td T$:
\begin{align}
    h\rightarrow \frac{6\sqrt{2}}{5l^3}\td T\,.
\end{align}
We see that using unimodular time $T$ as our evolution parameter is equivalent to gauge fixing the first approach using $N = \frac{6\sqrt{2}}{5l^3}$. Substituting this into (\ref{eq:lag5cell}) and converting the edge lengths to scale factors, we get
\begin{align}
    L(a,\dot{a}) = 3\left[\frac{\mathfrak{c}_1}{a^2}+\mathfrak{c}_2a^4\dot{a}^2\right]\,.
\end{align}
The coefficients are summarised in \cref{tab:compgflag} and compared to the continuous model. Again we see a slight preference for using $a_V$ over $a_R$.
\begin{table}[]
    \centering
    \begin{tabular}{c|c|c}
         & $\mathfrak{c}_1$ & $\mathfrak{c}_2$  \\\hline
         $a_V$ & $1.410$ & $746.470$ \\
         $a_R$ & $2.804$ & $94.964$ \\
         $a_{\text{cont.}}$ & $1$ & $4\pi^4\approx 389.636$ 
    \end{tabular}
    \caption{Comparison of our gauge-fixed Lagrangian in the continuous time limit to the gauge-fixed continuum FLRW model in (\ref{eq:FLRWgf}).}
    \label{tab:compgflag}
\end{table}
In this second version of taking the continuum limit we did not have to a priori assume that the height $h$ of the frusta is a good measure of time. Moreover, we could have done everything in terms of the edge lengths $m$ instead. In terms of unimodular time evolution the prescription gives
\begin{align}
    m \rightarrow \sqrt{\left(\frac{6\sqrt{2}}{5l^3}\right)^2+\frac{3}{8}\dot{l}^2}\,\td T\,.
    \label{eq:mcontinuum}
\end{align}
Plugging this into the action, calculating the continuous time Lagrangian and expanding it gives the same result as in (\ref{eq:lag5cell}) with $N = \frac{6\sqrt{2}}{5l^3}$. Thus, we have two different ways of arriving at the same continuum limit: one can calculate the height of the frusta, rewrite the action using this height instead of the edge length as the bulk variable and then take the continuum limit either using $h\rightarrow N\text{d}t$ or $\Delta T\rightarrow \text{d}T$. This procedure relies on symmetries of the bulk geometry and is a somewhat ad-hoc change of variables from the usual edge lengths to different ones. The second option is to stick with bulk edge lengths and use the unimodular condition and $\Delta T\rightarrow \text{d}T$ to define the continuum substitution (\ref{eq:mcontinuum}). In Henneaux--Teitelboim--Regge calculus the second option is more natural, as it retains the edge length as the bulk variable and unimodular time is a boundary variable that comes with the theory. For the model discussed in the next section, we will see that the definition of a bulk height variable is not possible, so the only way to compare it to the continuum is via the 4-volume. While the 4-volume can of course also be calculated in standard Regge calculus, in Henneaux--Teitelboim--Regge calculus it appears more naturally. Additionally, when considering Lorentzian Regge calculus using lengths as time becomes more subtle, as zero length does not necessarily correspond to no 4-volume between the boundaries (e.g., if the bulk edges are all null). Using 4-volume in the Lorentzian setting could alleviate some of these subtleties.

\subsection{Simplicial Model}
\label{sec:simmodel}

In this section we consider a simplicial complex $\mathcal{T}_{sim}$ whose boundary is again $\partial\mathcal{T}_{sim} = \partial\Delta^4\sqcup\partial\Delta^4$. As before, we will denote the vertices of the initial 5-cell by $\{x_i\}_{1\leq i\leq5}$ and the final vertices by $\{y_i\}_{1\leq i\leq5}$. Furthermore, we will denote a $p$-simplex by the number of $x$ and $y$ vertices it includes, e.g., a 2-simplex containing two $x$ vertices and one $y$ vertex is a triangle $t_{21}$.

The simplest way to connect the two 5-cells by a simplicial complex is to connect each vertex $x_i$ to all vertices in the final 5-cell apart from $y_i$. This means that for every tetrahedron in the initial 5-cell there is exactly one vertex $y_i$ in the final 5-cell that all vertices of the initial tetrahedron are connected to, forming a 4-simplex. 

In \cref{tab:numsim} we have summarised the relative numbers of each type of $p$-simplex; \cref{tab:tinsig} shows for each type of triangle which 4-simplices it is part of. These numbers are relevant for calculating the deficit angles for the Regge action.

\begin{table}[]
\centering
\begin{tabular}{|l|lllll|}
\hline
 \backslashbox{$\#x_j$}{$\#y_k$} & 0                          & 1                          & 2                          & 3                          & 4                         \\ \hline
0 & -                          & \cellcolor[HTML]{FFCCC9}5  & \cellcolor[HTML]{FFCE93}10 & \cellcolor[HTML]{FFFFC7}10 & \cellcolor[HTML]{9AFF99}5 \\
1 & \cellcolor[HTML]{FFCCC9}5  & \cellcolor[HTML]{FFCE93}20 & \cellcolor[HTML]{FFFFC7}30 & \cellcolor[HTML]{9AFF99}20 & \cellcolor[HTML]{CBCEFB}5 \\
2 & \cellcolor[HTML]{FFCE93}10 & \cellcolor[HTML]{FFFFC7}30 & \cellcolor[HTML]{9AFF99}30 & \cellcolor[HTML]{CBCEFB}10 & -                         \\
3 & \cellcolor[HTML]{FFFFC7}10 & \cellcolor[HTML]{9AFF99}20 & \cellcolor[HTML]{CBCEFB}10 & -                          & -                         \\
4 & \cellcolor[HTML]{9AFF99}5  & \cellcolor[HTML]{CBCEFB}5  & -                          & -                          & -                         \\ \hline
\end{tabular}
\caption{Number of simplices with $n$ $x$-vertices and $m$ $y$-vertices, where $n$ is given by the row and $m$ by the column. The colour indicates the dimension of the simplices: vertices (red), edges (orange), triangles (yellow), tetrahedra (green) or 4-simplices (blue).}
\label{tab:numsim}
\end{table}

\begin{table}[]
\centering
\begin{tabular}{l|llll}
         & $\sigma_{41}$ & $\sigma_{32}$ & $\sigma_{23}$ & $\sigma_{14}$ \\ \hline
$t_{30}$ & 2             & 1             & 0             & 0             \\
$t_{21}$ & 1             & 2             & 1             & 0             \\
$t_{12}$ & 0             & 1             & 2             & 1             \\
$t_{03}$ & 0             & 0             & 1             & 2            
\end{tabular}
\caption{This table shows in how many 4-simplices a triangle $t$ of a certain type is included and of what type they are.}
\label{tab:tinsig}
\end{table}

As before, we assume the edge lengths of the initial 5-cell are all the same and denoted by $l_1$. Similarly, we denote the edge lengths of the final 5-cell by $l_2$. The lengths of the bulk edges connecting an $x$ and a $y$ vertex are also assumed to be all equal and are denoted by $m$. Using this notation we can calculate the areas and 4-volumes of the simplices to be\footnote{The expressions for volumes of 4-simplices can be verified using, e.g., the Cayley--Menger determinant.}
\begin{align}
\label{eq:triareas}
    a_{30} &= \frac{\sqrt{3}l_1^2}{4}\,,\quad a_{03} = \frac{\sqrt{3}l_2^2}{4}\,,\quad
    a_{21} = \frac{l_1}{4}\sqrt{4m^2-l_1^2}\,,\quad a_{12} = \frac{l_2}{4}\sqrt{4m^2-l_2^2}\,,\\
    V_{41} &= \frac{\sqrt{2}l_1^3}{48}\sqrt{m^2-\frac{3}{8}l_1^2}\,,\quad V_{14} = \frac{\sqrt{2}l_2^3}{48}\sqrt{m^2-\frac{3}{8}l_2^2}\,,\\
    V_{32} &= \frac{l_1^2l_2}{16\sqrt{6}}\sqrt{2m^2-\frac{2}{3}l_1^2-\frac{1}{2}l_2^2}\,,\quad V_{23} = \frac{l_2^2l_1}{16\sqrt{6}}\sqrt{2m^2-\frac{2}{3}l_2^2-\frac{1}{2}l_1^2}\,.
\end{align}
To calculate the dihedral angles we proceed as for the cosmological model, starting with $\sigma_{41}$, which we embed into $\mathbb{R}^4$. To calculate the dihedral angle at the hinge $t_{30}$, we project everything to the plane orthogonal to $t_{30}$ and then calculate the 2-dimensional angle between the projected $\tau_{31}$ and $\tau_{40}$. Analogous steps are taken for $t_{21}$. The calculations result in the following values for the dihedral angles:
\begin{align}
    \theta^{41}_{30} = \cos^{-1}\left(\frac{1}{2}\frac{l_1}{\sqrt{6m^2-2l_1^2}}\right)\,,\quad \theta^{41}_{21} = \cos^{-1}\left(\frac{2m^2-l_1^2}{6m^2-2l_1^2}\right)\,.
\end{align}
From this we can immediately read off the expressions for the dihedral angles in $\sigma_{14}$:
\begin{align}
    \theta^{14}_{03} = \cos^{-1}\left(\frac{1}{2}\frac{l_2}{\sqrt{6m^2-2l_2^2}}\right)\,,\quad \theta^{14}_{12} = \cos^{-1}\left(\frac{2m^2-l_2^2}{6m^2-2l_2^2}\right)\,.
\end{align}
A simplex such as $\sigma_{41}$, built from a regular $(n-1)$-simplex and a vertex, called an apex, which has the same distance to all other vertices, is called a star-simplex. More general expressions for the dihedral angles of a star-simplex can be found for example in \cite{maehara2013dihedral}. 

Using the same method described above we calculate the dihedral angles of $\sigma_{32}$ to be
\begin{align}
\begin{split}
    \theta^{32}_{30} &= \cos^{-1}\left(\frac{3l_2^2-6m^2+2l_1^2}{2l_1^2-6m^2}\right)\,,\quad \theta^{32}_{21} = \cos^{-1}\left(\frac{l_1l_2}{2\sqrt{(3m^2-l_1^2)(4m^2-l_2^2-l_1^2)}}\right)\,,\\ \theta^{32}_{12} &= \cos^{-1}\left(\frac{2m^2-\frac{1}{2}l_2^2-l_1^2}{4m^2-l_2^2-l_1^2}\right)\,.
\end{split}
\end{align}
Analogously, for $\sigma_{23}$ we have
\begin{align}
\begin{split}
    \theta^{23}_{03} &= \cos^{-1}\left(\frac{3l_1^2-6m^2+2l_2^2}{2l_2^2-6m^2}\right)\,,\quad \theta^{23}_{12} = \cos^{-1}\left(\frac{l_1l_2}{2\sqrt{(3m^2-l_2^2)(4m^2-l_1^2-l_2^2)}}\right)\,,\\ \theta^{23}_{21} &= \cos^{-1}\left(\frac{2m^2-\frac{1}{2}l_1^2-l_2^2}{4m^2-l_1^2-l_2^2}\right)\,.
\end{split}
\end{align}
For this model, instead of looking at the action we consider directly the equations of motion. We can solve the constraint (\ref{eq:lambdacon}), so that we have one global constant $\Lambda$. Varying with respect to $m$ we get
\begin{align}
\label{eq:eomedge}
\begin{split}
        0 =& \,30\frac{\partial a_{t_{21}}}{\partial m}(2\pi-\theta^{41}_{21}-2\theta^{32}_{21}-\theta^{23}_{21})+30\frac{\partial a_{t_{12}}}{\partial m}(2\pi-\theta^{14}_{12}-2\theta^{23}_{12}-\theta^{32}_{12})\\ &-\Lambda\left(5\frac{\partial V_{41}}{\partial m}+10\frac{\partial V_{32}}{\partial m}+10\frac{\partial V_{23}}{\partial m}+5\frac{\partial V_{14}}{\partial m}\right)\,.
\end{split}
\end{align}
The complicated expressions for the angles make it hard to solve this equation explicitly, so we will need to resort to numerical methods.

\begin{figure}
    \centering
    \includegraphics[width=0.65\linewidth]{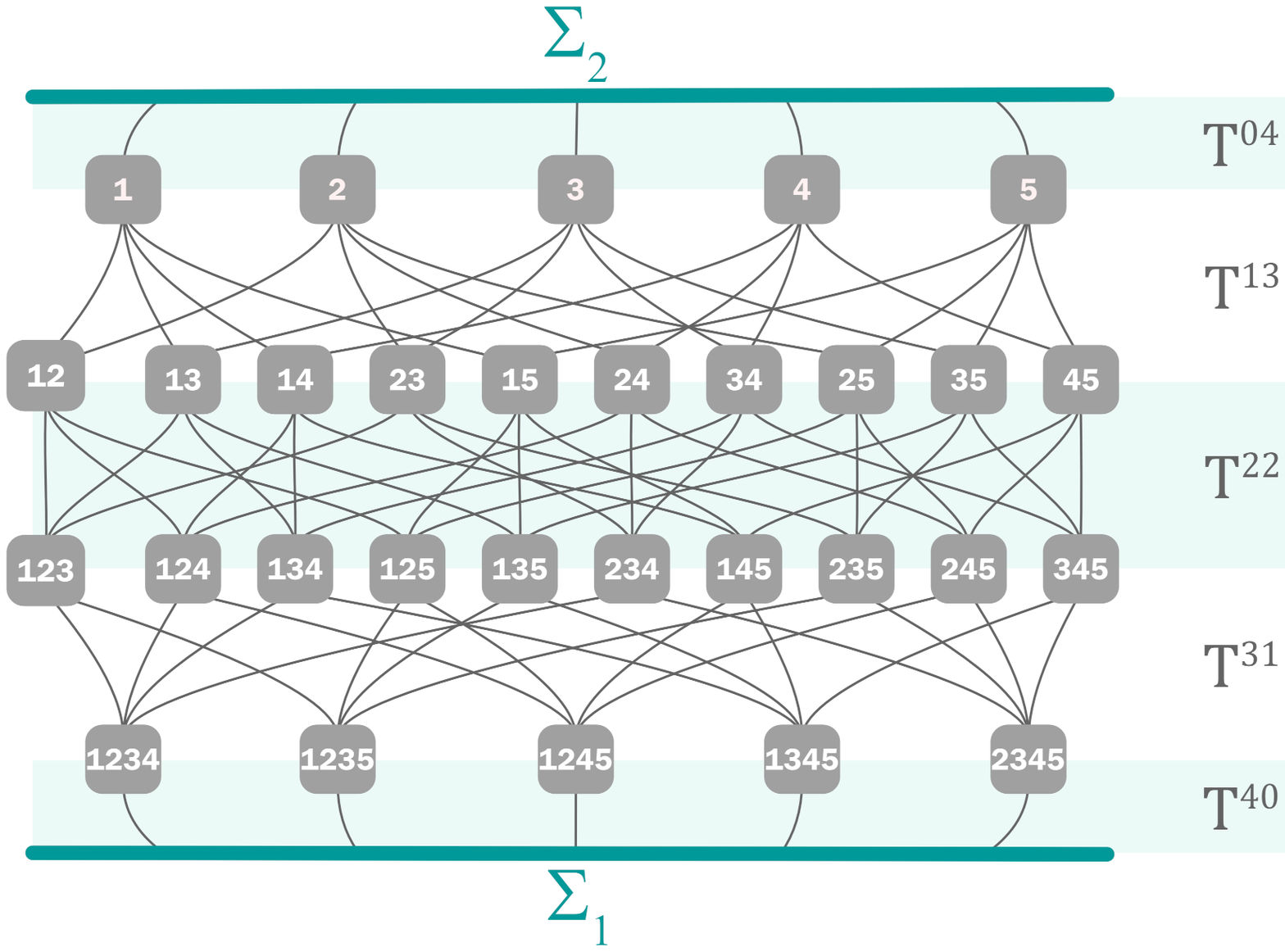}
    \caption{Visualisation of the unimodular evolution in $\mathcal{T}_{sim}$. The grey squares each symbolise a 4-simplex, the numbers on them denote which $x$ vertices are part of them, e.g., $1234$ is the 4-simplex made up of the vertices $x_1,\,x_2,\,x_3,\,x_4$ and $y_5$. The lines between the squares represent the tetrahedra which are each shared by two 4-simplices (except for the boundary tetrahedra). The boundary tetrahedra are represented by the lines connecting to either the initial hypersurface $\Sigma_1$ or the final hypersurface $\Sigma_2$. The background colours and the $T^{ij}$ depict the steps of unimodular evolution.}
    \label{fig:simevovis}
\end{figure}

The other equations of motion we need to solve are the unimodular condition for each 4-simplex, (\ref{eq:unicondis}). We assume that $T^{\tau}_{\sigma}$ is the same for all simplices of a given type. Under this assumption we have four different equations:
\begin{align}
    V_{41} &= T_{41}^{40}+4T_{41}^{31}\,,\quad V_{14} = T_{14}^{04}+4T_{14}^{13}\,,\\
    V_{32} &= 2T_{32}^{31}+3T_{32}^{22}\,,\quad V_{23} = 2T_{23}^{13}+3T_{23}^{22}\,.
\end{align}
In addition we have the conditions $T_{41}^{31} = -T_{32}^{31}$, $T_{14}^{13} = -T_{23}^{13}$ and $T_{32}^{22} = -T_{23}^{22}$ coming from (\ref{eq:tcon}). The value for $T_{41}^{40}$ is an initial condition we need to provide. Then given $l_1,l_2,m$ we get unique solutions for the $T^{\tau}_{\sigma}$:
\begin{align}
     T_{41}^{31} &= -T_{32}^{31} = \frac{1}{4}(V_{41}-T^{40}_{41})\,,\\
    T_{32}^{22} &= -T_{23}^{22} = \frac{1}{6}(V_{41}+2V_{32}-T^{40}_{41})\,,\\
     T_{23}^{13} &= -T_{14}^{13} =  \frac{1}{4}(V_{41}+2V_{32}+
    2V_{23}-T^{40}_{41})\,, \\
    T^{04}_{14} &= V_{41}+2V_{32}+2V_{23}+V_{14}-T^{40}_{41}\,.
\label{eq:T1}
\end{align}
In terms of unimodular time we find
\begin{align}
    T_2 = T_1 + 5V_{41}+10V_{32}+10V_{23}+5V_{14}\,,
\end{align}
which can be reformulated into $\Delta T = V_{\mathcal{T}_{sim}}$, as expected. Through the introduction of the unimodular time variables it is possible to answer the question ``How much time lies between the initial and final hypersurface?". In standard Regge calculus this question would be hard to answer, since there is no good proxy for time in a discrete simplicial complex. In a unimodular approach, one can fix $\Delta T$ as part of the boundary data and look for solutions with a fixed unimodular time elapsed between the boundaries. There are now also four different kinds of 4-dimensional building block: $\sigma_{41}$, $\sigma_{32}$, $\sigma_{23}$ and $\sigma_{14}$. The height of each of the 4-simplices is given by
\begin{align}
    h_{41/14} = \sqrt{m^2-\frac{3}{8}l_{1/2}^2}\,,\quad h_{32/23} = l_{2/1}\sqrt{\frac{m^2-\frac{1}{3}l_{1/2}^2-\frac{1}{4}l_{2/1}^2}{m^2-\frac{1}{3}l_{1/2}^2}}\,.
\end{align}
Because there is just a single free variable $m$ for the different bulk edge lengths, it is not possible for all of these heights to vanish simultaneously if $l_1,l_2\neq 0$. This makes it difficult to define a notion of time out of the geometry of the bulk and means that there is no straightforward continuum limit in which $\Delta T$ becomes infinitesimally small. The different types of 4-simplices can be ordered into four unimodular time steps as visualised in \cref{fig:simevovis}. 

\begin{figure}
     \centering
     \begin{subfigure}[b]{0.48\textwidth}
         \centering
         \includegraphics[width=\textwidth]{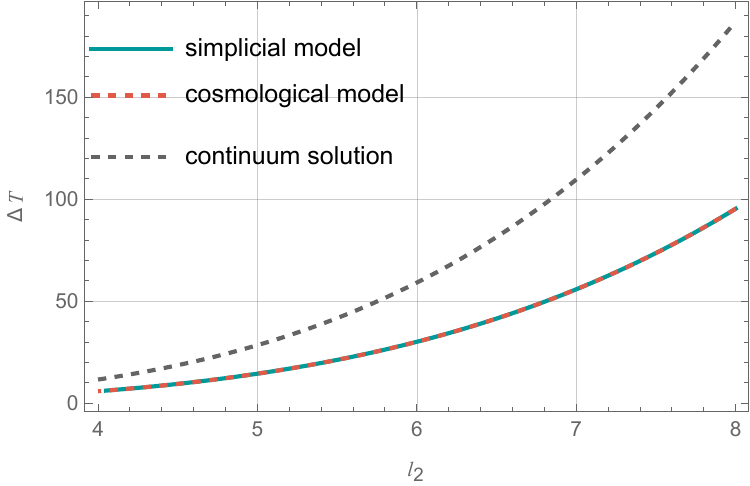}
         \caption{Plot for $\Lambda = 0$.}
     \end{subfigure}
     \hfill
      \begin{subfigure}[b]{0.48\textwidth}
         \centering
         \includegraphics[width=\textwidth]{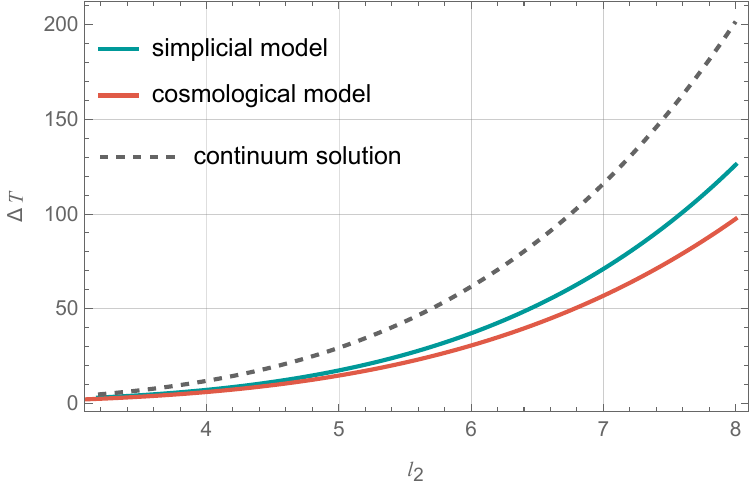}
         \caption{Plot for $\Lambda = 0.1$.}
     \end{subfigure}
     \hfill
     \begin{subfigure}[b]{0.48\textwidth}
         \centering
         \includegraphics[width=\textwidth]{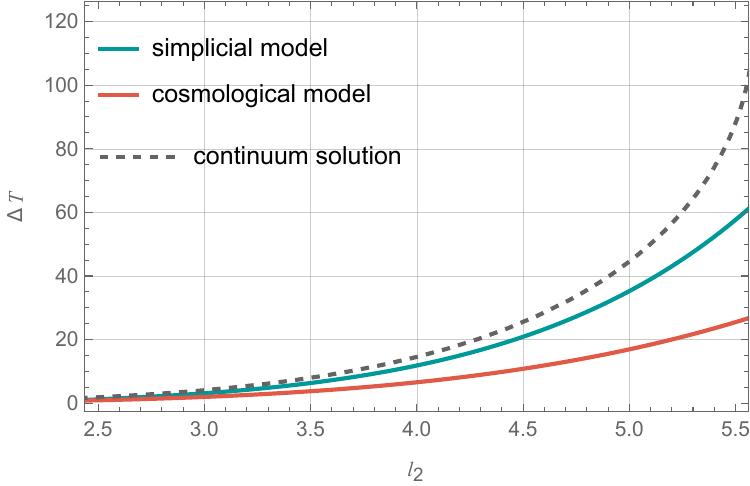}
         \caption{Plot for $\Lambda = 1$.}
     \end{subfigure}
        \caption{Accuracy of simplicial and cosmological models in approximating the continuum solution. In all plots $l_1 = 1$ and we have used the method of converting the edge length to a scale factor $a_V$ via the volume to compare to the continuum solution. The difference in unimodular time, i.e., the 4-volume between the hypersurfaces, is plotted as a function of $l_2$. Both the simplicial and the cosmological model clearly differ from the continuum solution, which is expected due to the coarse discretisation. The simplicial model agrees a bit better than the cosmological model for $\Lambda > 0$. In the case of $\Lambda = 0$ they are identical. For the simplicial model there is a value $ l_{\text{min}}$ for $l_2$ below which there may be no solution, the value of which depends on $\Lambda$. The plots only show $l_2 > l_{\text{min}}$.}
        \label{fig:eomedge}
\end{figure}

Even though there are no bulk vertices, the simplicial complex has a significantly more intricate structure compared to the cosmological model. Because the discretisation of the bulk is more complex, one might expect better agreement with the continuum solution. A comparison between the simplicial and cosmological model and the continuum solution discussed in \cref{minisupersec} is shown in \cref{fig:eomedge}. 
For both Regge models, the plots are computed numerically. For the plot of the continuum solution as a function of edge length, the volume conversion from scale factor to edge length, $a = a_V$ (see (\ref{eq:ltoa})), is used. If one instead uses the curvature conversion $a=a_R$, the mismatch between the continuum and the discrete models worsens. This fits our observation in the previous subsection that the volume conversion gives a result closer to the continuum.

In the case of $\Lambda = 0$, we see that the simplicial and cosmological models agree. The models differ significantly from the analytic continuum solution, which is to be expected, as the discretisation of a 3-sphere into five tetrahedra is a very coarse approximation. For the simplicial model there is a minimal value $l_{\text{min}} = 4l_1$ for $l_2$, below which no classical solutions to the Regge equations of motion exist. We will explain why this is the case in detail below. This fits in with our observation that there is no value of $m$ for which the heights of all 4-simplices vanish. It further confirms that a time continuum limit, such as the one calculated for the cosmological model, is not possible in this case, because the $l_2\rightarrow l_1$ limit is not well-defined.

In the case $\Lambda > 0$, the simplicial model agrees better with the continuum solution. There is again a minimal value for $l_2$ for the simplicial model, although in a slightly different sense. The difference is that in this case for some values $l_2 < l_{\text{min}}$ a solution to the equations of motion can still be found, but not for all $l_2 < l_{\text{min}}$. This could either be due to insufficient numerical analysis or to a fundamental limitation of the model.  We have plotted the solutions only in a range where we trust the numerics. We can observe the trend that increasing $\Lambda$ improves the agreement between the continuum solution and the simplicial model, whereas the agreement between the cosmological model and the continuum does not improve. Just as in the continuum, we see a maximum edge length (or scale factor) that depends on the value of the cosmological constant. In the continuum, the maximum is given by $a_{\text{max}} = \sqrt{\frac{3}{\Lambda}}$, which gives $l_{\text{max}} \approx 5.583$ for $\Lambda = 1$ using the volume conversion. In the simplicial model, we find $l_{\text{max}}\approx 8.626$, which is of the same order as the continuum result. In \cref{minisupersec}, we also saw that for $\Lambda > 0 $ there are two solutions for fixed boundary lengths (boundary scale factors), see (\ref{eq:solapm}). When solving the Regge equations numerically, we do not find both solutions; it seems that the discretisation eliminates one of the two solutions, although we have no definitive proof of this statement. The phenomenon of having a different number of solutions in the discrete and continuum cases was also noted in \cite{Dittrich:2021gww}, where Lorentzian solutions were observed in a regime where only Euclidean solutions should exist.

\begin{figure}
    \centering
    \includegraphics[width=0.75\linewidth]{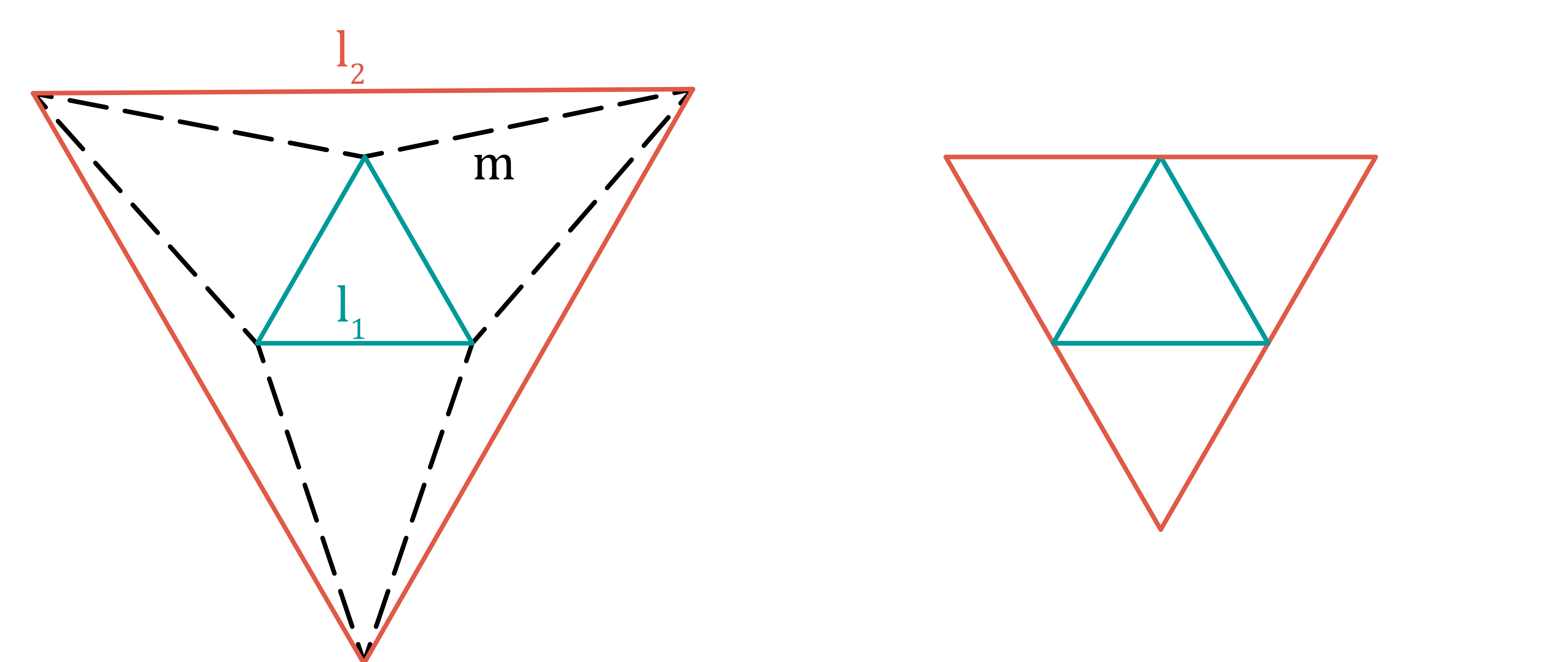}
    \caption{Visualisation of an analogue of the simplicial model in two dimensions. The bulk consists of two kinds of triangles, $t_{12}$ and $t_{21}$. For $l_2 = 2l_1$ the $t_{12}$ triangles become degenerate, and there are no classical solutions for $\frac{l_1}{2}<l_2<2l_1$.}
    \label{fig:minl22d}
\end{figure}

The reason there is no classical solution for $l_2<4l_1$ in the case of $\Lambda = 0$ can be understood when looking at the same construction in two dimensions (see \cref{fig:minl22d}). Since the continuum solution to general relativity is just flat space, we would expect the classical Regge solution to be close to a portion of flat space interpolating between the given boundary discretisations. In two dimensions, the bulk consists of two kinds of triangles, $t_{12}$ and $t_{21}$. If one decreases the length $l_2$ while keeping $l_1$ the same, there is a point at which the triangles $t_{12}$ become degenerate. This happens when $l_2 = 2l_1$; for smaller values of $l_2$ there is no flat solution.\footnote{If one decreases all the way to $l_2<\frac{l_1}{2}$ there are solutions again, mirroring those for $l_2>2l_1$ except with the final triangle inside the initial triangle.} Again, this observation shows clearly that the simplicial model has no continuum limit in which $l_1$ and $l_2$ are very close.

\begin{figure}
    \centering
    \includegraphics[width=0.75\linewidth]{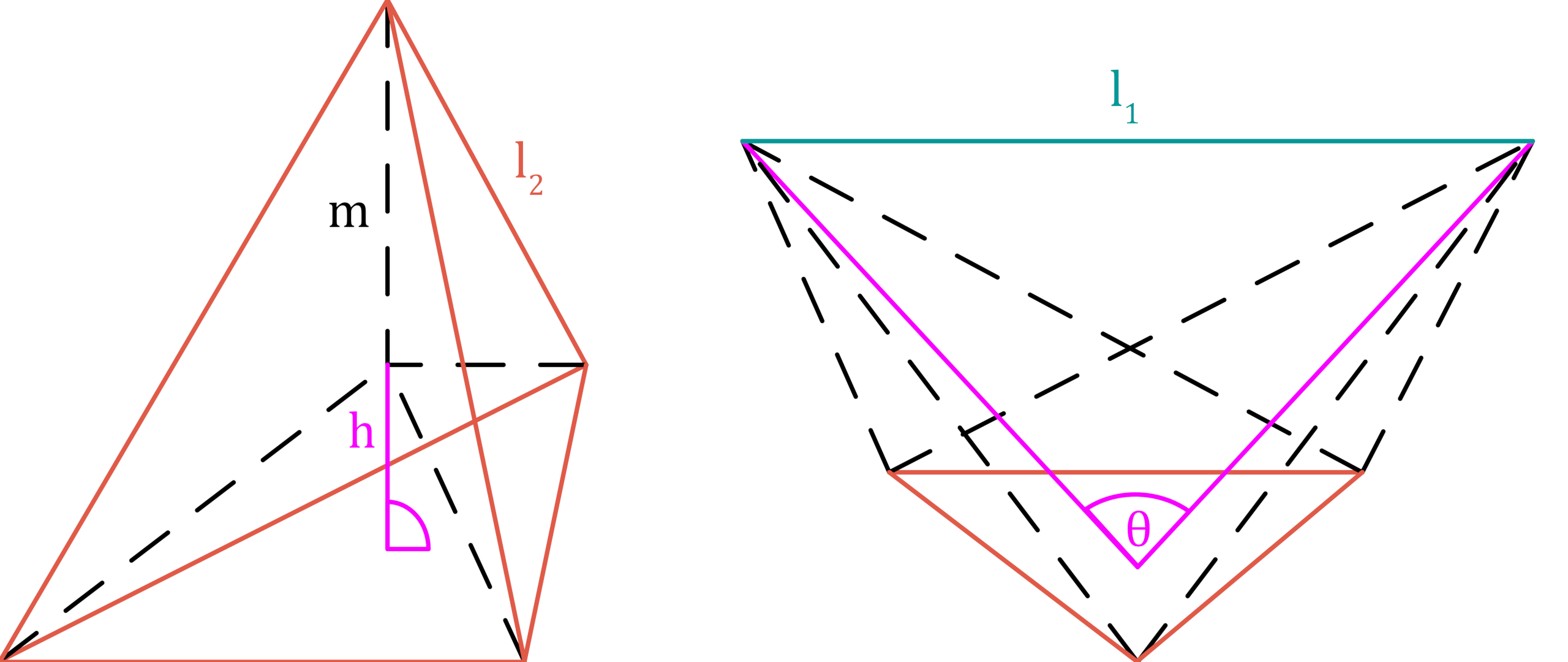}
    \caption{Left image: For $l_2 = l_{\text{min}}(l_1)$ the 4-simplices $\sigma_{14}$ become degenerate and $l_2$ can be calculated as a function of the height $h$. Right image: Using the dihedral angle of a regular 4-simplex $\theta$ the height $h$ can be related to the edge length $l_1$.}
    \label{fig:minl24d}
\end{figure}

\begin{figure}
    \centering
    \includegraphics[width=0.75\linewidth]{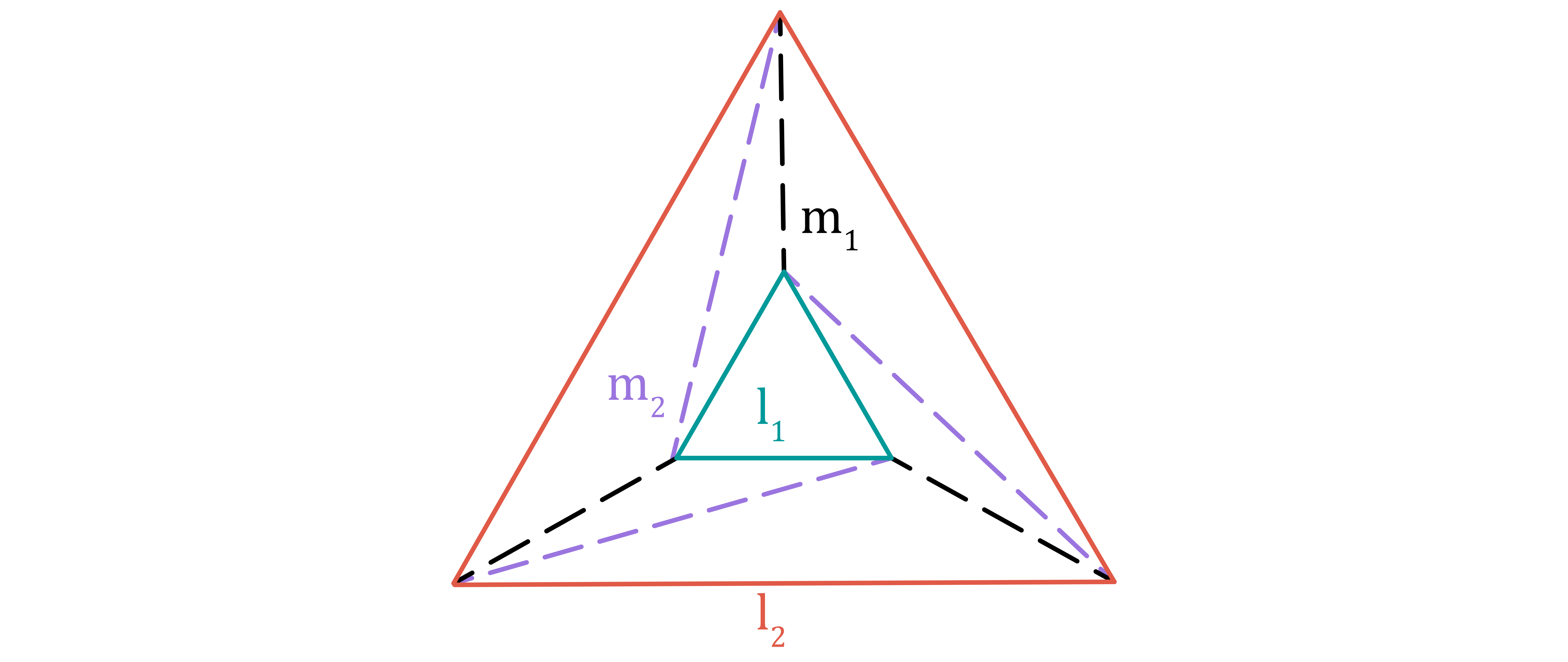}
    \caption{A two-dimensional simplicial model with two different edge lengths in the bulk, $m_1$ and $m_2$. Here there is no minimum value for $l_2$ for which a classical solution exists.}
    \label{fig:duall2d}
\end{figure}

Similarly, in four dimensions, for a given $l_1$, there is a minimal edge length $l_{\text{min}}(l_1)>l_1$ such that assuming $l_2>l_1$ actually requires $l_2 \ge l_{\text{min}}(l_1)$ for the existence of a classical solution. The value of $l_{\text{min}}$ can be calculated by considering the 4-simplex $\sigma_{14}$, which is the analogue of the triangle $t_{12}$ in two dimensions; the 4-simplex becomes degenerate when $l_2 = l_{\text{min}}(l_1)$ (see left side of \cref{fig:minl24d}). The height $h$ of a tetrahedron $\tau_{13}$ with base triangle $t_{03}$ in this case is given by $h = \frac{l_2}{2\sqrt{6}}$. On the other hand, we can consider the 4-simplex $\sigma_{23}$ (see right side of \cref{fig:minl24d}). We can also draw in the heights $h$ here, noting that the angle between the two heights of the tetrahedra of type $\tau_{13}$ is given by the dihedral angle of a regular 4-simplex, which is $\theta = \cos^{-1}(\frac{1}{4})$. Looking at the triangle made up of the two heights $h$ and the edge of length $l_1$, we can then see that
\begin{align}
    \sin\left(\frac{\theta}{2}\right) = \frac{l_1}{2h} = \frac{\sqrt{6}l_1}{l_2}\quad\Rightarrow\quad l_2 = 4l_1\,.
\end{align}
Thus, the simplicial model has the drawback that for $\Lambda = 0$ classical configurations with $\frac{l_1}{4}<l_2<4l_1$ are not possible. To include these we would need to allow for different lengths of the bulk edges; see \cref{fig:duall2d} for a different model in two dimensions that does allow for a continuum limit and has no such constraints.

We would expect similar constraints to be relevant for $\Lambda > 0$, even though the argument is more complicated given that the continuum solution would need to be embeddable into a 4-sphere, whose geometry is affected by our discretisation into simplices (unlike flat space for $\Lambda=0$). In the examples illustrated in \cref{fig:eomedge}, where we have chosen $l_1 = 1$, we do not find a hard cut-off as in the continuum. Instead, we find a value of $l_2 = l_{\rm min}$, below which there are still solutions for some values of $l_2$ but not all. Whether this is an artefact of the numerical limitations of our analysis or points to a more fundamental limitation of our model as in the case $\Lambda = 0$ cannot be decided definitively within our present analysis, and it is possible that the range can be extended. We leave exploring the small $l_2$ limit of the simplicial model for $\Lambda > 0$ in more detail to future work.

For the cosmological model, we demonstrated that unimodular time can be used to define the time continuum limit, analogously to the frustum height employed in previous work. The advantage of using unimodular time as the evolution parameter is that edge lengths are retained as fundamental bulk variables. We also considered a purely simplicial model demonstrating the applicability of unimodular Regge calculus in the simplicial setting. Although a time continuum limit cannot be defined in this setting, unimodular time still enables a direct comparison with the continuum. While standard Regge calculus could similarly utilise the 4-volume for continuum comparisons, unimodular Regge calculus holds a distinct advantage: the 4-volume arises naturally as the difference between boundary variables. Introducing $T^{\tau}_{\sigma}$ as boundary data allows us to consider time as something existing on a boundary hypersurface and not just in the evolution between hypersurfaces. This feature might help shed light on the nature of time in discrete quantum theories more generally, most obviously in quantum Regge calculus but also possibly in CDT and spin foams.

\section{Conclusion}
\label{sec:conclusion}

In this paper we have proposed a new Henneaux--Teitelboim--Regge action for discrete (Euclidean) gravity, defined in (\ref{eq:umreggeaction}). We have initiated its study by focusing on simple solutions that represent a homogeneous cosmological model and found that, just as in the continuum, the classical solutions of the unimodular version are mostly the same as the ones of standard Regge calculus. The differences are in the role of the cosmological constant $\Lambda$ in the theory and in the presence of the constraint that compact spacetimes without boundary must have zero 4-volume. A cosmological constant arising as a constant of integration is very appealing, as different values of $\Lambda$ correspond to different initial conditions in the same theory as opposed to distinct theories as in general relativity. An alternative angle on this feature is to say that Henneaux--Teitelboim unimodular gravity introduces a new degree of freedom called unimodular time $T$, which lives on the boundary of a triangulation and encodes the 4-volume between the boundaries. Hence, complete boundary data used to specify a spacetime history encodes the initial and final (unimodular) time, in contrast to general relativity where no boundary variable is directly interpretable as time.

We considered a simple model modelling the evolution of a discrete 3-sphere and were able to replicate the results in the literature on cosmological Regge calculus \cite{Liu:2015gpa,Dittrich:2021gww}. With regards to \cite{Dittrich:2021gww} this means deriving the time continuous limit of the Lagrangian and comparing it to the expression of continuum general relativity. We did this in several ways: first we rewrote the discrete action using the height of the 4-dimensional building blocks as a variable instead of the bulk edge length. This height variable can be identified with a time coordinate with arbitrary lapse $N$ in the time continuum limit, as in \cite{Dittrich:2021gww}. Alternatively, we used unimodular time to define a unique time variable in the continuum limit, which we saw is analogous to fixing the lapse to a particular value. Importantly, when using unimodular time to define the continuum limit, we do not need to rewrite the Regge action in terms of the height variables, but may use the bulk edge lengths used in Regge's original formulation. After expanding the Lagrangian to second order in velocity it can be compared to the continuum expression. To compare with \cite{Liu:2015gpa} we calculated the equations of motion and found that the expression for the Hamiltonian constraint agrees with the result of \cite{Liu:2015gpa} truncated to second order in velocity, while the time evolution (acceleration) equation is compatible with the Hamiltonian constraint in both cases. 

Furthermore, the addition of unimodular time allowed for a comparison to the continuum theory using boundary variables in a model with a complicated discrete structure. We considered a second model of the evolution of a discretised 3-sphere, keeping the boundary discretisation the same but with a more involved triangulation of the bulk. The main motivation for this model was that the discretisation is a simplicial complex, as in Regge's original formulation, and not a discretisation into generalised building blocks as in the first model. This model serves as a proof of concept that unimodular Regge calculus works in a simplicial setting without the need to impose symmetries of the bulk discretisation. For this model, no continuum limit in the sense described above can be defined, as the limit of the difference between initial and final edge lengths going to zero is not well-defined. Nonetheless, because of the additional degree of freedom of unimodular time, it is still possible to show that  for $\Lambda > 0$ the simplicial model approximates the continuum solution better in its range of validity than the first model we introduced. Since unimodular time is now part of the new boundary degrees of freedom in Henneaux--Teitelboim--Regge calculus, the use of unimodular time as a notion of time does not correspond to a gauge fixing. In general, the freedom to reparametrise the dynamics and choose a convenient time variable is not necessarily there in the discrete setting, since diffeomorphism invariance is broken (this is discussed for example in \cite{Dittrich:2008pw,Bahr:2009ku}). The boundary unimodular time gives us a diffeomorphism-invariant way of talking about time, which survives in the discrete setting. Furthermore, we found that increasing $\Lambda$ improves the agreement between the simplicial model and the continuum solution. Both models are based on the 5-cell as a relatively crude approximation to a continuum 3-sphere, and it would be worth investigating models based on more refined triangulations, again perhaps along the lines of \cite{Dittrich:2021gww}.

We showed that the solutions of unimodular Regge calculus are the same as for standard Regge calculus (with the caveats mentioned above). However, this would not be the case if we considered the corresponding quantum theory. Most obviously, a quantum unimodular Regge calculus would allow for superpositions of the value of $\Lambda$. As we mentioned in the introduction, we see Regge calculus as a good starting point for writing down a fundamental theory of unimodular quantum gravity, given that many approaches (e.g. quantum Regge calculus, CDT, spin foams and group field theory) are  related to the standard Regge action. It would be interesting to see how unimodular theories of quantum gravity differ from their standard counterpart. In asymptotic safety for example, the fixed point and condition for a UV complete Standard Model change significantly \cite{Eichhorn:2013xr,Eichhorn:2015bna,Benedetti:2015zsw,DeBrito:2019gdd,deBrito:2020rwu,deBrito:2020xhy}\footnote{In these papers the authors do not work with the Henneaux--Teitelboim formulation of unimodular gravity, but their results show that classically equivalent theories can differ in the quantum regime.}. It may thus well be that even though the theories are classically equivalent, their respective quantum theories differ considerably.

In the Henneaux--Teitelboim version of unimodular gravity specifically, the introduction of unimodular time $T$ might change the notion of time in a theory of quantum gravity and allow for questions to be asked that could not be asked in the usual ``timeless'' setting. Since $T$ is a variable on the boundary, it could potentially become an operator in a boundary Hilbert space, which would allow for time in a theory of quantum gravity to be studied in a novel way. 

Beyond quantisation, a natural extension of this work is the generalisation to Lorentzian signature. Defining the time continuum limit in Lorentzian Regge calculus is complicated by the presence of spacelike, timelike, and null edges. Formulating the framework with unimodular time could alleviate these difficulties, as the strictly positive 4-volume provides a well-defined, monotonic evolution parameter. There are also other formulations of Regge calculus, such as area Regge calculus \cite{Barrett:1997tx} and Regge calculus with constant curvature simplices \cite{Bahr:2009qd}. In principle it should be possible to make these theories unimodular in the same fashion as was done in this paper. 

\acknowledgments We thank Bianca Dittrich for insightful comments on the manuscript, and a referee for crucial comments on an earlier version. The work of SG was supported by the
Royal Society through the University Research Fellowship Renewal URF$\backslash$R$\backslash$221005. \Cref{fig:simplexvisual,fig:4simplex,fig:cosmobb,fig:simevovis,fig:minl22d,fig:minl24d,fig:duall2d} were made using Affinity by Canva.

\bibliographystyle{JHEP.bst}
\bibliography{unimodregge}

\end{document}